\documentclass[
 amsmath,amssymb,
 aps, physrev,
]{revtex4-2}

\usepackage{graphicx}
\usepackage{dcolumn}
\usepackage{bm}

\usepackage{algorithm}
\usepackage{algpseudocode}
\usepackage{booktabs} 
\usepackage[inkscape=true]{svg}
\usepackage{hyperref}

\definecolor{CB1}{rgb}{0.64, 0.0, 0.0}
\definecolor{CB2}{rgb}{0.03, 0.27, 0.49} 
\newcommand{\CB}[1]{{\color{black}{#1}}}

\begin{document}


\title{\textbf{Multi-objective Bayesian optimization of rigid and flexible nozzles for energy-efficient pulsed jet propulsion} 
}%




\author{Paras Singh}
\thanks{These authors contributed equally to this work.}
\affiliation{Daniel Guggenheim School of Aerospace Engineering, Georgia Institute of Technology, Atlanta, USA \relax}

\author{Yukesh Karki}
\thanks{These authors contributed equally to this work.}
\affiliation{Aerospace Engineering, School of Metallurgy and Materials, University of Birmingham, Birmingham, UK \relax}

\author{Victor Hernandez}
\affiliation{Daniel Guggenheim School of Aerospace Engineering, Georgia Institute of Technology, Atlanta, USA}

\author{Daehyun Choi}
\affiliation{School of Chemical and Biomolecular Engineering, Georgia Institute of Technology, Atlanta, USA \relax}

\author{Saad Bhamla}
\email{saad.bhamla@colorado.edu}
\affiliation{School of Chemical and Biomolecular Engineering, Georgia Institute of Technology}
\affiliation{BioFrontiers Institute, Department of Chemical and Biological Engineering, CU Boulder, Colorado, USA}

\author{Chandan Bose}
\email{c.bose@bham.ac.uk}
\affiliation{Aerospace Engineering, School of Metallurgy and Materials, University of Birmingham, Birmingham, UK}

\date{\today}

\begin{abstract}

The biomechanics of pulsed-jet propulsion in aquatic animals, including squids and jellyfish, provide valuable insights into energy-efficient locomotion. In these organisms, flexible funnel deformation enables rapid acceleration and maneuverability while minimizing energy expenditure. Drawing inspiration from these biological systems, this study investigates the performance trade-offs between rigid and flexible nozzle geometries in pulsed-jet propulsion systems. A multi-objective Bayesian optimization framework, integrated with three-dimensional fluid-structure interaction (FSI) simulations, is employed to identify nozzle designs that simultaneously maximize hydrodynamic impulse and minimize jet energy input. The optimization process reveals fundamentally distinct performance characteristics for rigid and flexible nozzles. Rigid nozzles achieve the highest impulse amplification, reaching up to 5 times the impulse produced by a baseline cylindrical nozzle, but this improvement is accompanied by substantially increased energy expenditure. In contrast, flexible nozzles yield lower peak impulse enhancement of approximately 2.5 times, while achieving significantly greater propulsion efficiency. The maximum normalized impulse-to-energy ratio attained by the flexible nozzles is approximately 1.8 times higher than that of rigid configurations, indicating a more effective conversion of input energy into useful propulsive output. Analysis of the underlying flow physics shows that optimized rigid nozzles enhance performance through geometry-induced internal entrainment, secondary vortex formation, and contraction-driven jet acceleration, resulting in stronger vortex circulation and downstream convection. Flexible nozzles utilize traveling expansion–contraction deformation waves that promote additional entrainment during expansion and accelerate the internally entrained fluid during contraction. These FSI-driven mechanisms improve pressure recovery, reduce pressure-energy expenditure, and mitigate negative pressure impulse contributions during the later stages of the pulse cycle. Present findings demonstrate that while rigid nozzle optimization is most effective for maximizing absolute impulse, flexible nozzles offer a more energy-efficient propulsion strategy for autonomous underwater vehicles.

\end{abstract}

\maketitle


\section{\label{sec:introduction}Introduction}

Pulsed jets are ubiquitous in nature, utilized by organisms, such as cephalopods, salps, and jellyfish, to achieve efficient underwater locomotion \cite{costello2021hydrodynamics,bujard2021resonant}. In contrast to steady continuous jets, pulsed jet propulsion expels a finite volume of fluid with high momentum into the surrounding medium \cite{krueger2003significance}. The propulsive thrust generated by these jets is primarily determined by the hydrodynamic impulse, which comprises both the jet's momentum flux and the pressure difference at the nozzle exit \cite{krieg2013modelling}. When a volume of fluid is impulsively accelerated through a nozzle, a shear layer is fed into the ambient fluid, which subsequently rolls up into a primary vortex ring \cite{didden1979formation, gharib1998universal}. The formation and pinch-off dynamics of this vortex ring are critical, as they govern the entrainment of ambient fluid and the resulting pressure variation, thereby influencing the total thrust generated by the system \cite{dabiri2009optimal, gao2020development}.

The thrust generated from traditional rigid nozzles is largely constrained by the piston stroke-to-diameter ratio of the jet generator and by the static geometric profiles \cite{gharib1998universal}. In contrast, the introduction of nozzle flexibility enables dynamic FSI effects that can substantially enhance the propulsive performance \cite{choi2022flow}. When subjected to a pulsatile jet, a flexible nozzle radially expands (as shown in Fig. \ref{fig:intro_schematic}) due to positive internal pressure, storing a portion of the work done by the fluid as elastic potential energy \cite{choi2024mechanism, choi2026squid, singh2026elastic}. This expansion initiates a traveling wave that propagates along the nozzle surface in accordance with Moens-Korteweg scaling \cite{choi2026squid, singh2026elastic}. The delayed expansion of the nozzle suppresses early shear-layer roll-up and delays the pinch-off of the primary vortex ring \cite{singh2026elastic}. Subsequently, as the nozzle contracts, the stored elastic energy is released, further accelerating the jet and significantly enhancing the exit velocity \cite{choi2022flow, choi2024mechanism}. This timely release of elastic energy increases both the total circulation and hydrodynamic impulse of the primary vortex ring, enabling the vortex ring to travel downstream at a higher translational velocity \cite{choi2022flow, choi2024mechanism}. For instance, compliant nozzles have been shown to increase primary vortex ring circulation by over 50\%, increase vortex convection distances by 9\%, and augment the total hydrodynamic impulse by over 60\% when compared to the rigid counterparts \cite{singh2026elastic}. Furthermore, during the abrupt deceleration phase of a pulsed jet, flexible nozzles exhibit an inward collapsing motion \cite{mitchell2025formation}. This collapse effectively suppresses the negative pressure gradient at the nozzle exit that typically reduces jet velocity and vortex circulation, leading to impulse increments of approximately 400\% and entrainment increments of 220\% \cite{choi2024mechanism}.

Despite the well-documented hydrodynamic benefits of FSI in pulsed jets, the majority of fundamental investigations have been restricted to uniform, cylindrical nozzle geometries \cite{choi2022flow, choi2024mechanism, mitchell2025formation}. The interplay of passive flexibility with non-cylindrical or complex geometric profiles remains largely unexplored. Because the wave propagation speed, energy storage, and associated vortex dynamics are highly sensitive to the spatial distribution of nozzle stiffness and internal flow geometry, deviating from cylindrical baseline designs presents an opportunity to uncover novel thrust-enhancing mechanisms. However, the design space for flexible propulsors, spanning diverse geometric profiles, lengths, and material stiffnesses, is highly non-linear. Assessing these complex, non-cylindrical designs requires computationally expensive three-dimensional FSI simulations \cite{morita2022applying} or time-consuming physical experiments \cite{choi2024mechanism}. Traditional gradient-based optimization methods are often ill-suited for such systems due to the absence of gradient information and their tendency to get trapped in local optima within complex, non-linear, and non-convex design space topologies. Therefore, robust and data-efficient optimization strategies are required to effectively navigate this extensive design space.

Bayesian Optimization (BO) provides a data-efficient framework for the global optimization of expensive, derivative-free black-box functions, which makes it particularly suitable for complex FSI problems \cite{frazier2018tutorial}. BO constructs a probabilistic surrogate model, most commonly using Gaussian Process Regression (GPR), to approximate the true objective function from a limited set of prior evaluations and to rigorously quantify predictive uncertainty \cite{morita2022applying}. An acquisition function, such as Expected Improvement (EI), is subsequently used to determine the next evaluation point by systematically balancing the exploitation of known high-performing regions with the exploration of areas characterized by high uncertainty \cite{frazier2018tutorial}. Although standard Bayesian Optimization (BO) is highly effective for single-objective problems, the design of rigid and flexible propulsor nozzles inherently involves competing physical requirements. Focusing solely on hydrodynamic impulse without constraining jet input energy does not accurately represent the actual design constraints. This approach may yield designs that achieve high impulse but require significantly greater energy inputs, which in turn reduces overall propulsive efficiency. Consequently, optimizing these nozzles necessitates balancing trade-offs among multiple conflicting objectives, a challenge well addressed by Multi-Objective Bayesian Optimization (MOBO) \cite{konakovic2020diversity, galuzio2020mobopt}.

Unlike single-objective methods that converge to a scalar solution, MOBO seeks to identify a set of Pareto-optimal designs, defined as configurations in which improvement in one objective necessitates compromise in another \cite{konakovic2020diversity}. To address these competing physical requirements, the present study utilizes MOBO to systematically investigate the non-cylindrical design space. This approach enables the identification of optimal rigid and flexible nozzle geometries that maximize hydrodynamic impulse while minimizing the required jet energy input. To reduce the dimensionality of the parametric design space and facilitate efficient exploration, geometric profiles are constrained to be axisymmetric, and a uniform material formulation is applied to all flexible designs. This framework offers a systematic pathway for developing highly efficient nozzle designs for pulsed-jet-based underwater propulsion.

The primary objectives of this study are three-folds: (i) to integrate a MOBO framework with the high-fidelity three-dimensional FSI simulations to identify Pareto-optimal rigid and flexible nozzle geometries that maximize hydrodynamic impulse and minimize jet energy input at the same time; (ii) to quantify the propulsive trade-offs between rigid and flexible nozzle classes through parallel optimization campaigns conducted under identical flow conditions, establishing a direct comparison of their impulse-to-energy characteristics; and (iii) to examine the physical mechanisms governing the performance of the Pareto-optimal designs through vortex dynamics and impulse-energy analyses, with the aim of elucidating the distinct roles of geometry-induced internal entrainment in rigid nozzles and FSI-driven pressure recovery in flexible nozzles. The results reveal that, in rigid nozzles, geometry-induced internal entrainment and secondary vortex formation dominate. In flexible nozzles, traveling expansion-contraction waves with pressure-recovery effects are the governing mechanisms.

The remainder of the paper is organized as follows. The computational domain, governing equations, solver setup, and validation studies are described in \S\,\ref{sec:methodology}, along with the nozzle parameterization and the MOBO framework in \S\,\ref{sec:optimization_framework}. The optimization convergence and Pareto-front analysis, followed by an examination of the vortex dynamics and FSI mechanisms of the Pareto-optimal nozzle designs, and a detailed impulse and energy analysis are presented in \S\,\ref{sec:results}. Finally, salient findings of this study are summarized, and conclusions are drawn in \S\,\ref{sec:conclusions}.

\section{\label{sec:methodology}Computational Methodology}

\subsection{Problem Definition}

\begin{figure}[htbp]
	\centering
	\includegraphics[width=\columnwidth]{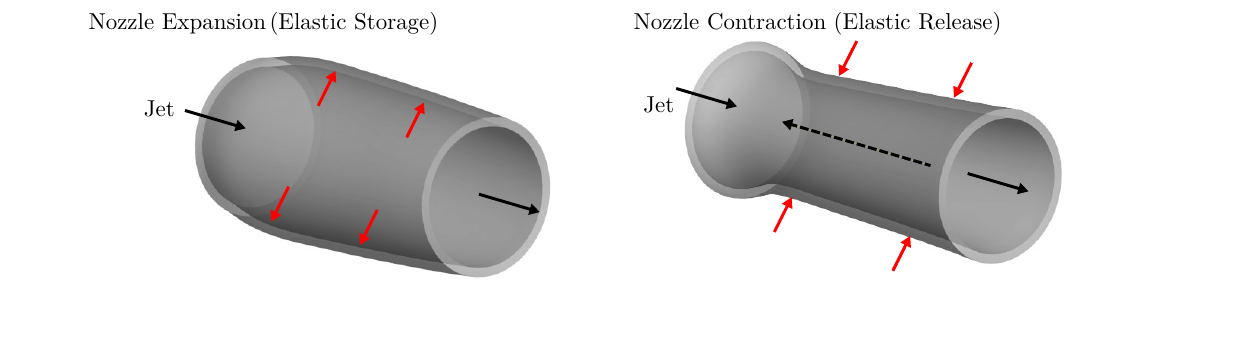}
	\caption{Schematic representation of a cylindrical nozzle system with flexible walls that passively expand and contract in response to internal pulsatile flow. The compliant walls store elastic potential energy during expansion and release it during contraction to enhance the jet, thus augmenting thrust.} 
	\label{fig:intro_schematic}
\end{figure}

In this study, the thrust generation and vortex impulse characteristics of rigid and flexible nozzles with varying geometries, when subjected to a pulsatile jet, are investigated. All nozzle configurations share a fixed inlet internal diameter of $D = 15~\mathrm{mm}$, ensuring identical inflow conditions across all designs. The nozzle length $L$ is treated as a design parameter and is discussed in subsequent sections. A cylindrical nozzle with $L = 40~\mathrm{mm}$ is selected as the baseline configuration following the work of Choi and Park \cite{choi2024mechanism}. \CB{The present simulations are performed under the same pulsed-jet conditions reported in the experiments of Choi and Park \cite{choi2024mechanism}. The prescribed inlet velocity profile is shown in Fig. \ref{fig:verification_validation}(a), where the jet reaches a peak velocity of $u_{\mathrm{jet}} = 0.293~\mathrm{m\,s^{-1}}$ after an acceleration period of $T_{\mathrm{acc}} = 0.05~\mathrm{s}$. The corresponding Reynolds number, defined as $Re = u_{\mathrm{jet}}D/\nu$, is approximately $4395$ based on the nozzle inlet diameter and the peak jet velocity. The flexible nozzle is modeled as a silicone-rubber structure with density $\rho_s = 1300~\mathrm{kg\,m^{-3}}$, uniform wall thickness $h = 1~\mathrm{mm}$, and Poisson's ratio $\nu_s = 0.4$. The present work employs a fixed Young's modulus of $E = 100~\mathrm{kPa}$ corresponding to a nozzle stiffness $Eh=100~\mathrm{N\,m^{-1}}$.} Figure~\ref{fig:intro_schematic} illustrates the operating principle of a flexible cylindrical nozzle subjected to internal pulsatile flow. During the acceleration phase, the positive internal pressure drives radial expansion of the compliant nozzle wall, storing a portion of the fluid work as elastic potential energy. As the flow decelerates, the stored elastic energy is released through nozzle contraction, accelerating the jet and augmenting the thrust generated at the nozzle exit.

\subsection{Governing Equations}

The fluid flow is governed by the incompressible laminar Navier-Stokes equations in the arbitrary Lagrangian-Eulerian (ALE) form:
\begin{equation}
\label{continuity}
\nabla \cdot \mathbf{u}=0,
\end{equation}
\begin{equation}
\label{momentum}
\frac{\partial\mathbf{u}}{\partial t}+ [(\mathbf{u}-{\mathbf{u}}^m) \cdot \nabla]\mathbf{u}=
-\frac{\nabla p}{\rho} + \nu\nabla^2\mathbf{u}.
\end{equation}
Here, $\mathbf{u}$ is the flow velocity, $\mathbf{u}^m$ is the grid point velocity, p is fluid pressure, and $\rho$ is fluid density. 

The structural behavior of the elastic nozzle is governed by the linear momentum balance equation,
\begin{equation}
    \rho_s \frac{\partial^2 \mathbf{d}_s}{\partial t^2} - \nabla \cdot \boldsymbol{\sigma}_s = 0,
\end{equation}
where $\rho_s$ denotes the solid density, $\mathbf{d}_s$ is the displacement field, and $\boldsymbol{\sigma}_s$ represents the Cauchy stress tensor.

Assuming a linear elastic material, the stress-strain relationship is given by
\begin{equation}
    \boldsymbol{\sigma}_s = \lambda \, \mathrm{tr}(\boldsymbol{\varepsilon}) \mathbf{I} + 2\mu \boldsymbol{\varepsilon},
\end{equation}
where $\lambda$ and $\mu$ are the Lamé parameters, corresponding to the first Lamé constant and the shear modulus, respectively, and $\mathbf{I}$ is the identity tensor. The infinitesimal strain tensor $\boldsymbol{\varepsilon}$ is defined as
\begin{equation}
    \boldsymbol{\varepsilon} = \frac{1}{2} \left( \nabla \mathbf{d}_s + (\nabla \mathbf{d}_s)^{T} \right).
\end{equation}

\CB{\subsection{Computational Domain, Mesh, and Boundary Conditions}}

\begin{figure}[htbp]
    \centering
    \includegraphics[width=\linewidth]{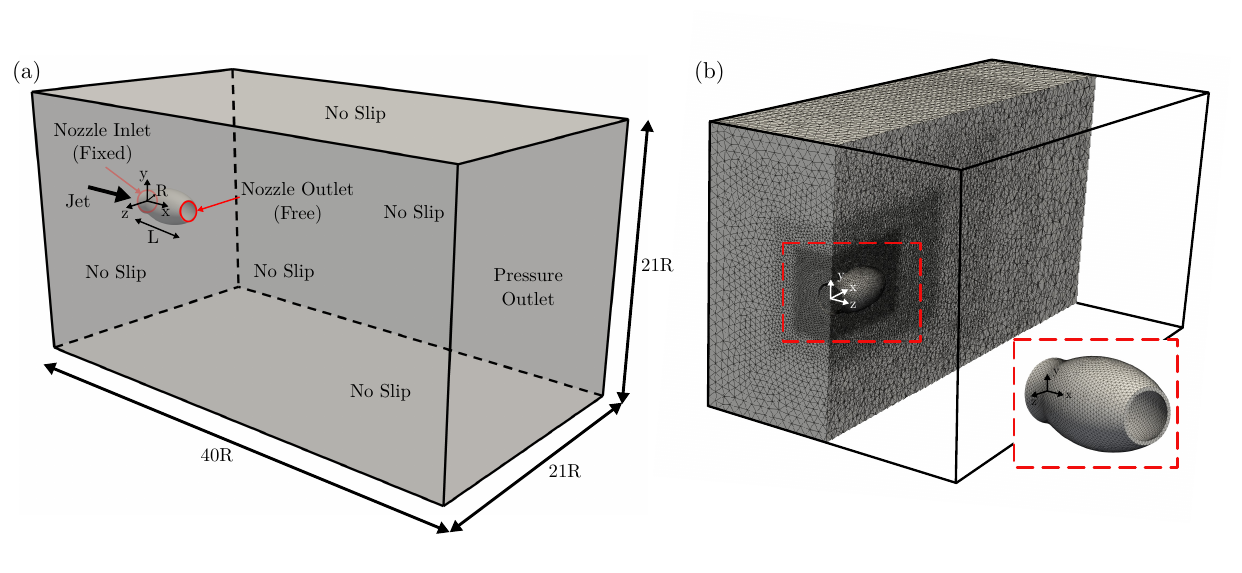}
    \caption{(a) Schematic \CB{representation of the computational domain and details of} boundary conditions employed for the coupled fluid and solid solvers. (b) \CB{Unstructured tetrahedral meshes used for the fluid and solid domains with local refinement in the vicinity of the nozzle.} The inset provides a detailed view of the nozzle surface mesh.}
    \label{fig:schematic}
\end{figure}

The computational domain \CB{used for the present simulations and the} associated boundary conditions are illustrated in Fig. \ref{fig:schematic}(a). A three-dimensional rectangular flow domain of dimensions $40R \times 21R \times 21R$ ($R$ is the nozzle inlet radius) is employed, where the flow field is described in the Cartesian coordinate system $(x, y, z)$. The $x$-axis is aligned with the nozzle centerline, and the origin is located at the center of the nozzle inlet. The domain dimensions were selected to ensure that the developing jet remains sufficiently distant from the external boundaries, thereby minimizing confinement effects and artificial boundary interactions for all nozzle geometries considered during the optimization study.

\CB{Figure} \ref{fig:schematic}(b) shows the computational mesh used for both the fluid and solid domains. Unstructured tetrahedral elements are employed throughout the simulation to accurately represent complex nozzle geometries and to maintain mesh quality during large structural deformations. To reduce computational cost while preserving solution accuracy, local mesh refinement is applied in the vicinity of the nozzle and within the near-field jet region, where large velocity gradients and FSI are expected. The mesh is gradually coarsened away from these regions to maintain a smooth size transition. Additionally, comparable element sizes are maintained across the fluid–solid interface to improve the accuracy of force transfer and displacement mapping between the two solvers.

At the nozzle inlet, a prescribed jet velocity condition is imposed such that $\mathbf{u}=\mathbf{u}_{\mathrm{in}}$, where $\mathbf{u}_{\mathrm{in}}$ denotes the specified inflow velocity profile presented in Fig. \ref{fig:verification_validation}(a). No-slip boundary conditions are enforced on all solid surfaces, including the nozzle wall and the outer boundaries of the computational domain, resulting in $\mathbf{u}=\mathbf{0}$. At the downstream boundary, a pressure-outlet condition is applied by prescribing the ambient pressure, $p=p_{\infty}$.

For simulations involving flexible nozzles, the nozzle inlet is fully constrained (clamped) to represent rigid attachment to the mounting surface, while the remainder of the nozzle, including the outlet, is free to deform under fluid loading as shown in Fig. \ref{fig:schematic}(a). Two-way FSI coupling is enforced at the fluid--solid interface. The kinematic continuity condition requires the fluid velocity at the interface to match the structural velocity, i.e., $\mathbf{u}=\partial \mathbf{d}_s/\partial t$, where $\mathbf{d}_s$ is the structural displacement vector. In addition, traction equilibrium is imposed across the interface through the condition $\boldsymbol{\sigma}_f \cdot \mathbf{n}_f + \boldsymbol{\sigma}_s \cdot \mathbf{n}_s = \mathbf{0}$, where $\boldsymbol{\sigma}_f$ and $\boldsymbol{\sigma}_s$ denote the fluid and solid Cauchy stress tensors, respectively, and $\mathbf{n}_f$ and $\mathbf{n}_s$ are the corresponding outward unit normal vectors. Together, these conditions ensure the consistent transfer of deformation and \CB{fluid forces} between the fluid and structural domains throughout the simulation.\\

\subsection{Solver Setup}

The governing fluid flow equations are solved using the finite volume method implemented in \CB{the open-source code {\tt OpenFOAM}} \cite{weller1998tensorial}. A transient formulation is employed, with temporal discretization based on a first-order implicit Euler scheme. Spatial discretization of the gradient terms is performed using a cell-limited Gauss linear scheme to ensure numerical stability, while maintaining second-order accuracy in smooth regions. The divergence term in the momentum equation is discretized using a linear upwind scheme, which provides a good balance between accuracy and boundedness. Diffusion terms are discretized using a second-order Gauss linear scheme with limiter, and linear interpolation is used for face values. Surface-normal gradients are computed using a limited scheme to control numerical oscillations. Pressure-velocity coupling is achieved using the {\tt PIMPLE} algorithm, which combines features of both  {\tt PISO} (Pressure-Implicit with Splitting of Operators) and {\tt SIMPLE} (Semi-Implicit Method for Pressure-Linked Equations) methods. In the present study, five outer corrector loops are employed, with three inner corrector steps and three non-orthogonal correction steps per time step to ensure stability during mesh motion. The pressure equation is solved using a geometric-algebraic multigrid (GAMG) solver with Gauss-Seidel smoothing for intermediate iterations, while the final pressure correction is computed using a preconditioned conjugate gradient (PCG) solver. The velocity field is solved using a stabilized bi-conjugate gradient (PBiCGStab) solver with a diagonal incomplete LU (DILU) preconditioner. A strict convergence tolerance of $10^{-8}$ is imposed for the velocity solver, while the pressure solver uses a tolerance of $10^{-6}$ with a relative tolerance of $10^{-2}$ for intermediate iterations and zero relative tolerance for final corrections. For mesh motion associated with the flexible nozzle, the structural displacement equation is solved using a PCG solver with a tolerance of $10^{-8}$.

The structural response of the flexible nozzle is computed using \CB{the open-source code {\tt CalculiX}} \cite{dhondt2017calculix}, which employs an implicit finite element formulation. A transient dynamic analysis is performed to accurately capture the transient deformation and inertial effects induced by the fluid forces. Time integration is handled via the Hilber-Hughes-Taylor $\alpha$-method \cite{miranda1989improved}, where the numerical damping parameter $\alpha$ governs the level of high-frequency dissipation in the solution. An appropriate value of $\alpha$ is selected to ensure numerical stability while preserving the physically relevant structural response. Geometric nonlinearity is taken into account by enabling large deformation effects. The chosen solver settings provide a stable and robust framework for the FSI simulations.

\begin{figure}[h!]
    \centering
    \includegraphics[width=\linewidth]{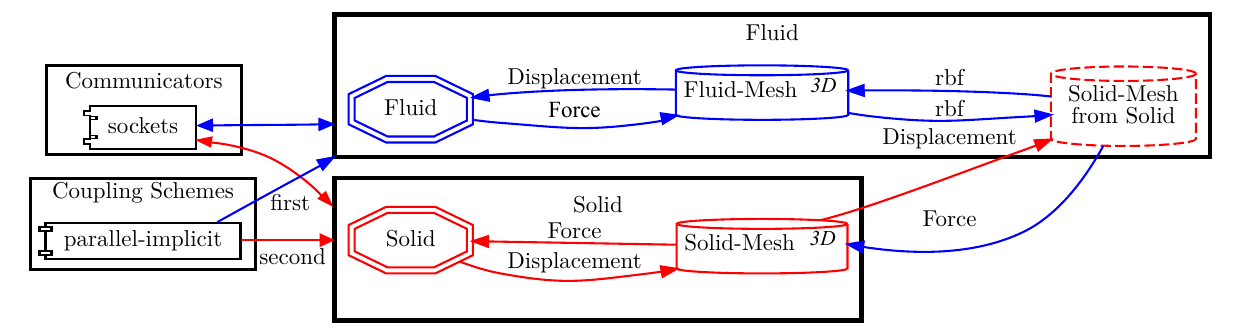}
    \caption{Graphical representation of fluid-structure solver coupling between {\tt OpenFOAM} (fluid solver) and {\tt CalculiX} (structural solver) using {\tt preCICE} (FSI coupling library).}
    \label{fig:preciceConfig}
\end{figure}

The fluid and structural solvers are coupled using the partitioned multi-physics coupling library {\tt preCICE} \cite{chourdakis2023openfoam}, as shown in Fig. \ref{fig:preciceConfig}. A Dirichlet-Neumann coupling strategy is adopted, where structural displacements are transferred to the fluid solver, and fluid forces are communicated back to the structural solver at the interface. The coupling is performed in a strongly coupled manner within each time step to ensure stability and accuracy of the FSI. A radial basis function (RBF) mapping scheme is employed to transfer data between the non-matching fluid and structural meshes at the interface. To accelerate the convergence of the coupled system, the interface coupling iterations are stabilized using the Interface Quasi-Newton Inverse Least Squares (IQN-ILS) method. The simulations are advanced in time using a fixed time step size, $\Delta t = 5\times10^{-4}~\mathrm{s}$, and coupling iterations are performed within each time step until a prescribed convergence criterion based on the interface residuals is satisfied.

\subsection{Verification and Validation Studies}

\CB{To assess the influence of spatial and temporal discretization on the numerical solution, mesh and time-step convergence studies were performed using the Richardson extrapolation procedure \cite{celik2008procedure}. The results are summarized in Table~\ref{table_mesh_time_step_Richardson_extrapolation}. For the mesh convergence study, the maximum normalized nozzle-exit velocity, $(u/u_{\mathrm{jet}})_{\max}$, was evaluated for both rigid and flexible ($Eh=150~\mathrm{N\,m^{-1}}$) baseline cylindrical nozzle configurations using three successively refined meshes. The coarse, medium, and fine meshes (Mesh 1, 2, and 3, respectively) contained approximately $0.5\times10^6$, $1.0\times10^6$, and $2.0\times10^6$ cells, respectively. Temporal convergence was assessed using the medium mesh while varying the time-step size over three levels of refinement. The estimated discretization errors and Richardson-extrapolated solutions indicate satisfactory convergence with respect to both mesh density and time-step size. Consequently, the medium-resolution mesh containing approximately $1.0\times10^6$ cells and a time step of $\Delta t = 5\times10^{-4}~\mathrm{s}$ was selected for all simulations, providing a suitable balance between accuracy and computational cost.}

\begin{table}[htbp]
\centering
\caption{Mesh and time-step convergence study based on the maximum normalized nozzle-exit velocity, $(u/u_{\mathrm{jet}})_{\max}$. Richardson extrapolation is used to estimate the asymptotic solution and discretization errors.}
\label{table_mesh_time_step_Richardson_extrapolation}

\begin{tabular}{lccccccc}
\toprule
\textbf{Case} &
\multicolumn{3}{c}{\textbf{Mesh Convergence $(u/u_{\mathrm{jet}})_{\max}$}} &
\textbf{Extrapolated} &
$\mathbf{e_{12}}$ &
$\mathbf{e_{23}}$ &
$\mathbf{e^{\mathrm{extr}}}$ \\

\cmidrule(lr){2-4}

&
\textbf{Mesh 1} &
\textbf{Mesh 2} &
\textbf{Mesh 3} &
\textbf{Solution} &
\textbf{[\%]} &
\textbf{[\%]} &
\textbf{[\%]} \\

\midrule

Rigid nozzle
& 0.849 & 0.857 & 0.857
& 0.857
& 0.920 & 0.033 & 0.001 \\

Flexible nozzle
& 1.393 & 1.402 & 1.405
& 1.406
& 0.680 & 0.205 & 0.089 \\

\midrule

\textbf{Case} &
$\boldsymbol{\Delta t = 1\times10^{-3}\,\mathrm{s}}$ &
$\boldsymbol{\Delta t = 5\times10^{-4}\,\mathrm{s}}$ &
$\boldsymbol{\Delta t = 2.5\times10^{-4}\,\mathrm{s}}$ &
\textbf{Extrapolated} &
$\mathbf{e_{12}}$ &
$\mathbf{e_{23}}$ &
$\mathbf{e^{\mathrm{extr}}}$ \\

\midrule

Rigid nozzle
& 0.851 & 0.857 & 0.858
& 0.858
& 0.659 & 0.090 & 0.014 \\

Flexible nozzle
& 1.355 & 1.402 & 1.413
& 1.416
& 3.370 & 0.755 & 0.220 \\

\bottomrule
\end{tabular}
\end{table}

\begin{figure}[htbp]
    \centering
    \includegraphics[width=\linewidth]{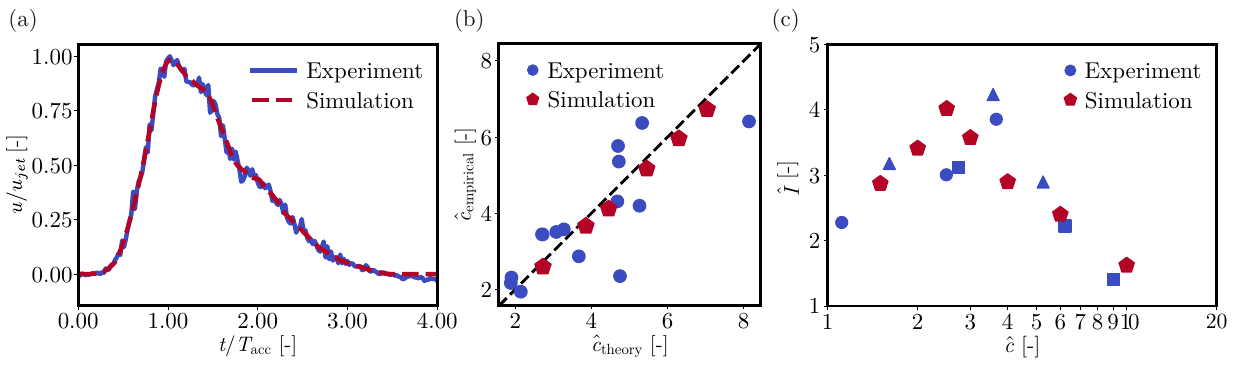}
    \caption{(a) Prescribed inlet pulsed jet velocity profile normalized by $u_{jet}$ and $T_{\mathrm{acc}}$. (b) Validation of the simulated non-dimensional wave speed against theoretical predictions and experimental measurements. (c) Normalized hydrodynamic impulse generated across different wave speeds, highlighting optimal propulsive condition between $\hat{c} \in [2, 4]$ for cylindrical nozzles. Experimental benchmark data adapted from Choi and Park \cite{choi2024mechanism}.}
    \label{fig:verification_validation}
\end{figure}

The validation of the coupled FSI simulation framework utilized in this study has been extensively established in our prior investigations \cite{choi2026squid, singh2026elastic}, wherein the computational results were compared against the experimental data of Choi and Park \cite{choi2024mechanism}. To briefly highlight this validation, the dynamic structural response of the baseline cylindrical nozzle is evaluated by examining the propagation of elastic waves along its surface. The theoretical wave propagation speed ($\hat{c}_{theory}$) is dictated by the non-dimensional Moens-Korteweg wave speed, formulated as $\hat{c} = \sqrt{E h T_{\mathrm{acc}}^2 / (\rho_f D L^2)}$, where $\rho_f$ denotes the density of the working fluid. By tracking the spatio-temporal evolution of the nozzle wall during the initial radial expansion cycle, the experimental and numerical non-dimensional wave speeds ($\hat{c}_{empirical}$) are extracted from the nozzle deflection data \cite{choi2026squid}. As presented in Fig. \ref{fig:verification_validation}(b), the computational predictions generated by our numerical framework exhibit close agreement with both the analytical Moens-Korteweg scaling and the experimental measurements provided by \cite{choi2024mechanism}. 

\CB{Furthermore}, to ensure the FSI solver accurately computes the propulsive performance metrics critical to our optimization objectives, the computational framework is validated against the optimal impulse condition for cylindrical nozzles reported experimentally by Choi and Park \cite{choi2024mechanism}. The total hydrodynamic impulse generated by the jet, $I$, is quantified as the sum of the momentum impulse ($I_m$) and the pressure impulse ($I_p$) \cite{gao2020development} and is given by Eq. \ref{eq:impulse}. As established previously, the non-dimensional wave speed $\hat{c}$ governs the FSI dynamics and is dependent on the jet inlet profile. To systematically evaluate the impulse response across a range of wave speeds, the jet acceleration time $T_{acc}$ is varied while maintaining fixed geometric dimensions and a constant nozzle stiffness of $Eh = 75~\mathrm{N\,m^{-1}}$ for this specific validation subset. To ensure kinematic consistency between the acceleration and deceleration phases across all variations of $T_{acc}$, the inlet axial velocity profile is parameterized as a continuous harmonic function:
\begin{equation}
u_f(t) = 
\begin{cases} 
\tfrac{u_{jet}}{2}\!\left[ 1 - \cos\!\left( \tfrac{\pi t}{T_{\mathrm{acc}}} \right) \right], & 0 \le t < T_{\mathrm{acc}}. \\[6pt]
\tfrac{u_{jet}}{2}\!\left[ 1 + \cos\!\left( \tfrac{\pi (t - T_{\mathrm{acc}})}{T_{\mathrm{acc}}} \right) \right], & T_{\mathrm{acc}} \le t \le 2T_{\mathrm{acc}}.
\end{cases}
\end{equation}
The total hydrodynamic impulse for each flexible configuration is subsequently normalized ($\hat{I}$) by the impulse generated by a rigid nozzle subjected to an identical $T_{\mathrm{acc}}$ jet profile. As illustrated in Fig. \ref{fig:verification_validation}(c), the normalized impulse magnitudes computed via our numerical framework exhibit qualitative trends that are consistent with the experimental data. Most importantly, the solver successfully captures the optimal wave speed condition, predicting peak impulse enhancement within the range of $\hat{c} \in [2, 4]$.

It should be noted that this validation is not intended to yield an exact quantitative replication of the experimental impulse magnitudes. Rather, the primary objective is to demonstrate the solver's ability in capturing the optimal wave speed threshold, which shows its effectiveness in capturing the complex FSI dynamics inherent to flexible nozzles. The minor deviations in absolute magnitude are expected; the reference experiments evaluated nozzles with extreme flexibility ($Eh < 45~\mathrm{N\,m^{-1}}$), a physical condition that is difficult to replicate in ALE-based FSI simulations due to the onset of severe mesh distortion. Therefore, accurately capturing the governing physical trends and the optimal kinematic scaling provides a robust validation of the computational framework and confirms the solver's capability to evaluate propulsive efficiency for the subsequent MOBO studies.

\section{\label{sec:optimization_framework}Optimization Framework}
\CB{\subsection{Design variables and objective functions}}

\begin{figure}[h]
    \centering
    \includegraphics[width=\linewidth]{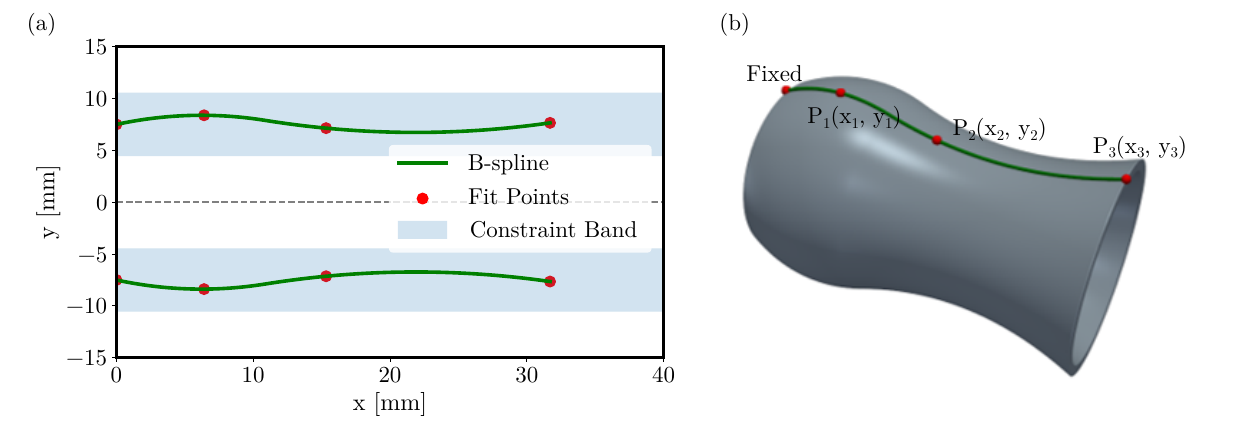}
    \caption{(a) Two-dimensional nozzle geometry showing the spline control points and the highlighted admissible region within which the control points are allowed to vary during optimization; (b) corresponding three-dimensional axisymmetric configuration.}
    \label{fig:parameterization}
\end{figure}

The nozzle geometry is parameterized using a two-dimensional B-spline representation of the nozzle profile, which is subsequently revolved about the centerline to generate the three-dimensional axisymmetric configuration. The geometry is described using six design variables,

\begin{equation}
\mathbf{x} =
[x_1,\, x_2,\, x_3,\, y_1,\, y_2,\, y_3]^T
\in \chi \subset \mathbb{R}^6,
\end{equation}

\noindent where \(x_1\), \(x_2\), and \(x_3\) denote the axial locations of the spline control points, while \(y_1\), \(y_2\), and \(y_3\) define their relative radial location with respect to the baseline cylindrical configuration. The admissible bounds for the design variables are summarized in Table~\ref{tab:bounds}. Fig. \ref{fig:parameterization}(a) illustrates the spline-based parameterization of the nozzle contour using three control points $P_1$, $P_2$, and $P_3$. The associated three-dimensional axisymmetric nozzle geometry obtained through revolution of the spline profile is shown in Fig.~\ref{fig:parameterization}(b). 

The inlet diameter is maintained fixed throughout the optimization process, whereas both the nozzle length and outlet diameter vary depending on the location of the terminal control point \(P_3\). The upper bound of \(x_3 = 40~\mathrm{mm}\) corresponds to the cylindrical nozzle length of the baseline configuration, thereby ensuring that all candidate geometries remain within a practically relevant design space.

\begin{table}[h]
\centering
\caption{Design variable bounds for the spline-based nozzle parameterization.}
\begin{tabular}{c c}
\hline
\textbf{Variable} & \textbf{Constraints} [mm] \\
\hline
$x_1$ & $[3,\;13]$ \\
$x_2$ & $[x_1+2,\;26]$ \\
$x_3$ & $[x_2+2,\;40]$ \\
$y_1$, $y_2$, $y_3$ & $[-3,\;3]$ \\
\hline
\end{tabular}
\label{tab:bounds}
\end{table}

To ensure smooth and physically realizable nozzle geometries, additional geometric constraints are imposed on the control points. In particular, a minimum axial spacing of \(2~\mathrm{mm}\) is enforced between consecutive control points, such that \(x_{i+1} - x_i \ge 2~\mathrm{mm}\) for \(i = 1,2\), thereby preventing spline overlap, excessive curvature, and nonphysical nozzle shapes. Furthermore, the radial perturbations are restricted within prescribed bounds to maintain manufacturable geometries and avoid abrupt area variations along the nozzle wall. Additionally, these parameter bounds provide sufficient freedom for geometric variation while maintaining a compact and computationally tractable design space.

The propulsive performance of the nozzles is evaluated using two quantities, the outlet impulse, denoted by \(I(\mathbf{x})\), and the inlet energy, denoted by \(E(\mathbf{x})\). The outlet impulse represents the total momentum flux and pressure thrust generated at the nozzle exit over the actuation period as give by \cite{gao2020development}:

\begin{equation}
I = \int_{0}^{t}
\int_{A_{\mathrm{out}}}
\left[
\rho u_x^2(\mathbf{r},\tau)
+
\left(p(\mathbf{r},\tau)-p_{\infty}\right)
\right]
\, dA \, d\tau,
\label{eq:impulse}
\end{equation}

\noindent where \(u_x\) and \(p\) denote the axial velocity and pressure at the nozzle outlet, respectively, \(\rho\) is the fluid density, \(p_{\infty}\) is the ambient pressure, and \(A_{\mathrm{out}}\) represents the outlet cross-sectional area.

The energy input at the inlet is evaluated through the time-integrated energy flux entering the nozzle,

\begin{equation}
E =
\int_{0}^{t}
\int_{A_{\mathrm{in}}}
\left(
p\,u_x
+
\frac{1}{2}\rho u_x^3
\right)
\, dA \, d\tau,
\label{eq:energy}
\end{equation}

\noindent where \(A_{\mathrm{in}}\) denotes the inlet cross-sectional area. The first term in Eq.~(\ref{eq:energy}) represents the pressure work rate, while the second term corresponds to the kinetic energy flux associated with the incoming flow.

The problem is formulated as a bi-objective optimization task that aims to maximize the generated impulse while simultaneously minimizing the inlet energy expenditure:

\begin{equation}
\max_{\mathbf{x} \in \chi}
\left(
I(\mathbf{x}),
- E(\mathbf{x})
\right).
\end{equation}

The evaluation of objective functions for each nozzle designs require a high-fidelity numerical simulation, which is computationally expensive. Furthermore, in the absence of gradient information, the mapping from the design variables to the objective space, \(\mathbf{x} \mapsto \left(I(\mathbf{x}), E(\mathbf{x})\right)\), is treated as a deterministic black-box function. Under these conditions, conventional gradient-based optimization methods become impractical. To efficiently explore the design space while minimizing the number of expensive function evaluations, the optimization problem is addressed using a multi-objective Bayesian optimization framework.

\CB{\subsection{Bayesian optimization}}

\begin{figure}[h]
    \centering
    \includegraphics[width=\linewidth]{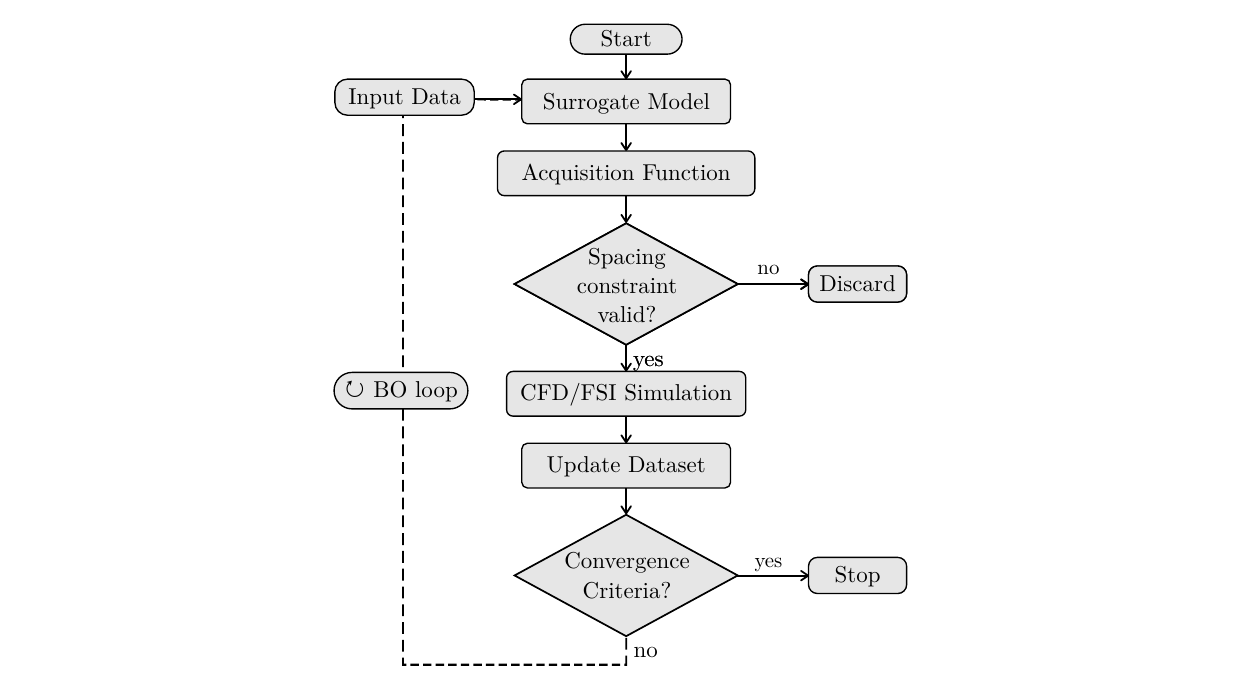}
    \caption{Multi-objective Bayesian optimization framework}
    \label{fig:flowchart}
\end{figure}

The Bayesian optimization (BO) framework is summarized in Fig. \ref{fig:flowchart}. An initial set of design points is generated using Latin Hypercube Sampling (LHS), a statistical sampling technique that ensures a well-distributed coverage of the design space. LHS divides each design variable range into equally probable intervals and samples each interval exactly once, thereby reducing clustering and improving space-filling properties in high-dimensional domains. In the present study, twenty initial design points are generated using LHS to provide a representative exploration of the six-dimensional parameter space. These samples serve as the initial training dataset for constructing the surrogate models and enable the Bayesian optimization process to begin with a diverse and informative set of observations.

Due to the high computational cost of evaluating the physics-based simulations at each design point, Gaussian process (GP) surrogate models are employed to approximate the response surface. GP regression provides both a predictive mean, $\mu(\mathbf{x})$, and an associated uncertainty, $\sigma(\mathbf{x})$, making it particularly suitable for Bayesian optimization. Two independent single-task GP models are constructed using the {\tt ModelListGP} framework in BoTorch~\cite{balandat2020botorch, ament2023unexpected}, corresponding to the normalized impulse $I/I_0$ and normalized energy $E/E_0$, where $I_0$ and $E_0$ are the values for the baseline cylindrical nozzle. Each GP model employs a Mat\'{e}rn-5/2 kernel with 
automatic relevance determination (ARD) to capture spatial correlations 
between the design variables. The kernel hyperparameters, including the 
length-scale vector $\boldsymbol{\ell}$ and noise variance $\sigma_n^2$, 
are estimated by maximizing the marginal log-likelihood:

\begin{equation}
\hat{\boldsymbol{\theta}} =
    \arg\max_{\boldsymbol{\theta}}
    \sum_{k=1}^{2}
    \log p(\mathbf{y}_k \mid \mathbf{X}, \boldsymbol{\theta}_k),
\end{equation}

\noindent where $\mathbf{X}$ denotes the training input matrix, $\mathbf{y}_k$ 
represents the corresponding training outputs for objective $k$, and 
$\boldsymbol{\theta}_k$ contains the kernel and likelihood parameters 
of the $k^{th}$ GP model. Optimization of the marginal likelihood is 
performed using the limited-memory Broyden--Fletcher--Goldfarb--Shanno 
algorithm with bound constraints (L-BFGS-B) via the 
\texttt{fit\_gpytorch\_mll} routine. To conform with BoTorch's 
maximization convention, the energy objective is negated, resulting in 
the transformed training objectives 
$\mathbf{y}_{\mathrm{train}} =
[I/I_0,\; -E/E_0]$.

New candidate designs are selected by maximizing the $q$-batch Log Expected Hypervolume Improvement (\texttt{qLogEHVI}) acquisition function~\cite{ament2023unexpected, daulton2020differentiable}. This acquisition function estimates the expected improvement in dominated hypervolume upon evaluating a batch of $q$ candidate points:

\begin{equation}
    \alpha(\mathbf{X}_q) =
    \mathbb{E}\left[
        \log\left(
            1 + \mathrm{HVI}(\mathbf{P} \cup \mathbf{X}_q, \mathbf{r})
            - \mathrm{HVI}(\mathbf{P}, \mathbf{r})
        \right)
    \right],
\end{equation}

\noindent where the hypervolume improvement is defined as

\begin{equation}
\Delta \mathrm{HVI}(\mathbf{X}_q)
=
\mathrm{HVI}(\mathbf{P}\cup\mathbf{X}_q,\mathbf{r})
-
\mathrm{HVI}(\mathbf{P},\mathbf{r}),
\end{equation}

\noindent where $\mathbf{P}$ denotes the current Pareto front and $\mathbf{r}$ is a reference point defined as the component-wise minimum of observed objective values minus a small offset ($0.1$). The logarithmic transformation mitigates vanishing gradients in regions far from the Pareto front, improving optimization robustness.

The expectation is approximated using a Monte Carlo estimator with $N=128$ Sobol samples:

\begin{equation}
\alpha(\mathbf{X}_q) 
\approx
\frac{1}{N}
\sum_{i=1}^{N}
\log
\left(
1+\Delta \mathrm{HVI}(\mathbf{X_q},\omega_i)
\right),
\end{equation}

\noindent where $\omega_i$ denotes the $i^{th}$ sample drawn from the joint GP posterior distribution. The hypervolume improvement $\Delta \mathrm{HVI}(\mathbf{X_q},\omega_i)$ represents the increase in dominated hypervolume obtained by adding the sampled objective realization of the candidate batch $\mathbf{X_q}$ to the current Pareto front. Differentiability is ensured via the reparameterization:

\begin{equation}
f^*(\mathbf{x})=\mu(\mathbf{x})+L(\mathbf{x})\epsilon,
\qquad
\epsilon \sim \mathcal{N}(0,I),
\end{equation}

\noindent where $f^*(\mathbf{x})$ represents a sample from the joint GP posterior, $\mu(\mathbf{x})$ is the posterior mean prediction, $L(\mathbf{x})$ is the Cholesky factor of the posterior covariance matrix, and $\epsilon$ is a standard normal random vector. This reparameterization transforms the stochastic GP sampling operation into a differentiable mapping with respect to the candidate design variables. Consequently, gradients of the \texttt{qLogEHVI} acquisition function can be efficiently computed through automatic differentiation, enabling gradient-based optimization of the acquisition function.

The optimization proceeds iteratively over multiple generations. At each generation, GP models are refitted using all available data, and the acquisition function is maximized over the bounded design space. To avoid local optima, a multi-start strategy is used. Specifically, 128 Sobol initializations are generated, from which the top 10 are selected via Boltzmann sampling and refined using L-BFGS-B. Optimization terminates when convergence criteria are satisfied or a maximum of 100 iterations is reached. A batch size of $q=5$ is used. Geometric feasibility is enforced through spacing constraints:
\begin{equation}
    x_2 > x_1 + 2,
    \quad
    x_3 > x_2 + 2.
\end{equation}

Valid candidate designs are evaluated using high-fidelity simulations, and the resulting data are appended to the dataset:
\begin{equation}
    \mathcal{D} \leftarrow \mathcal{D} \cup
    \{\mathbf{X}^*, f(\mathbf{X}^*)\}.
\end{equation}

The iterative optimization process is considered to have converged when the relative change in the hypervolume (HV) indicator between successive iterations falls below a prescribed tolerance, i.e.\ $\left|HV_i - HV_{i-1}\right| / HV_{i-1} < \varepsilon$, where $HV_i$ denotes the hypervolume at generation $i$, and $\varepsilon$ is a user-defined convergence threshold. The HV quantifies the objective space dominated by a set of Pareto-optimal solutions relative to a chosen reference point \cite{yang2019efficient}. For a Pareto approximation set ($\mathcal{P} = {\mathbf{y}^{(1)}, \ldots, \mathbf{y}^{(n)}} \subset \mathbb{R}^{d}$), the HV is expressed as
\begin{equation}
    HV(\mathcal{P}) = \lambda_d \left( \bigcup_{\mathbf{y} \in \mathcal{P}} [\mathbf{r}, \mathbf{y}] \right),
\end{equation}
where $\lambda_d$ denotes the $d$-dimensional Lebesgue measure and $\mathbf{r}$ represents the reference point. This metric evaluates the size of the dominated region enclosed between the Pareto front and the reference point. The multi-objective Bayesian optimization framework is summarized in Fig. \ref{fig:flowchart} and Algorithm \ref{alg:algorithm}. Further details on the MOBO algorithm can be found in \cite{galuzio2020mobopt, balandat2020botorch, ament2023unexpected}.

\begin{algorithm}[H]
\caption{Multi-Objective Bayesian Optimization Framework}
\label{alg:algorithm}
\begin{algorithmic}[1]

\State \textbf{Input:} Initial dataset $\mathcal{D} = \{\mathbf{X}, \mathbf{y}\}$

\State Initialize hypervolume metric $HV_0$

\For{iteration $i = 1, 2, \dots$}

\State \textbf{Step 1:} Fit GP surrogate models by maximizing the marginal log-likelihood:
\State \hspace{1em} $\hat{\boldsymbol{\theta}} = \arg\max_{\boldsymbol{\theta}} \sum_{k=1}^{2} \log p(\mathbf{y}_k \mid \mathbf{X}, \theta_k)$

\State \textbf{Step 2:} Construct Pareto front $P$ and define reference point $\mathbf{r}$

\State \textbf{Step 3:} Define acquisition function $\alpha(\mathbf{X}_q)$ using qLogEHVI with frozen Sobol samples

\State \textbf{Step 4:} Optimize acquisition function using multi-start L-BFGS-B:
\For{each restart $j = 1, \dots, N_{\text{restart}}$}
    \State Initialize $\mathbf{X}^{(0)}_j$
    \For{$t = 1, \dots, T_{\max}$}
        \State Compute gradient $\nabla \alpha(\mathbf{X}^{(t)}_j)$
        \State Update $\mathbf{X}^{(t+1)}_j$ using L-BFGS-B
        \If{$\|\nabla \alpha\| < \tau_g$ \textbf{or} $|\Delta \alpha|/\max(|\alpha|,1) < \tau_f$ \textbf{or} $t = T_{\max}$}
            \State \textbf{break}
        \EndIf
    \EndFor
\EndFor
\State Select best candidate $\mathbf{X}^*$

\State \textbf{Step 5:} Enforce spacing constraints:
\If{$x_2 > x_1 + 2$ \textbf{and} $x_3 > x_2 + 2$}
    \State Accept candidate
\Else
    \State Reject candidate
    \State \textbf{continue}
\EndIf

\State \textbf{Step 6:} Evaluate accepted candidates using CFD simulation
\State Update dataset: $\mathcal{D} \leftarrow \mathcal{D} \cup \{\mathbf{X}^*, f(\mathbf{X}^*)\}$

\State Compute hypervolume $HV_i$

\If{$\left|HV_i - HV_{i-1}\right| / HV_{i-1} < \varepsilon$}
    \State \textbf{break} \Comment{Convergence criterion met}
\EndIf

\EndFor

\State \textbf{Output:} Final Pareto-optimal design set

\end{algorithmic}
\end{algorithm}

\section{Results and Discussion}
\label{sec:results}
\subsection{Optimization Convergence and Pareto Front Analysis}
\label{subsec:convergence}

The convergence behavior of the MOBO framework is evaluated using the hypervolume (HV) indicator. The HV represents the extent of the objective space covered by the Pareto-optimal solutions relative to a specified reference point. Consequently, the progression of HV over successive function evaluations provides an effective measure of optimization performance, capturing both the convergence of the solutions toward the true Pareto front and the diversity of their distribution across the trade-off space.
    
\begin{figure}[h!]
    \centering
    \includegraphics[width=\linewidth]{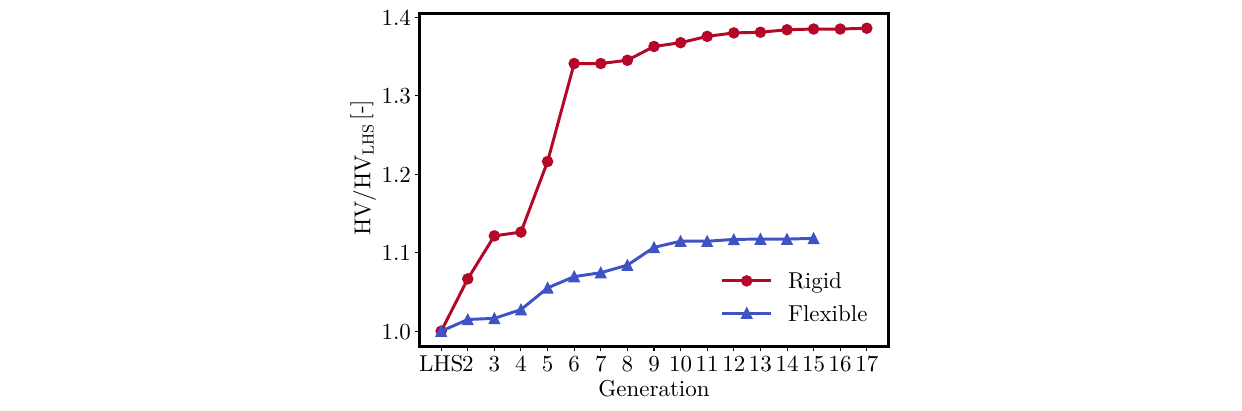}
    \caption{Normalized hypervolume (HV) as a function of MOBO generations for rigid and flexible nozzle configurations. The HV is normalized by the initial LHS population and serves as a measure of Pareto-front quality and diversity. Both cases exhibit rapid early improvement followed by saturation, indicating convergence of the optimization process.}
    \label{fig:convergence}
\end{figure}

For comparison between the rigid and flexible nozzle optimizations, the HV is normalized against the initial value obtained from the Latin Hypercube Sampling (LHS) dataset. As illustrated in Fig. \ref{fig:convergence}, the normalized HV grows significantly during the initial iterations, indicating rapid progress in identifying improved Pareto-optimal solutions across the objective space. After approximately 6 MOBO iterations for the rigid nozzles and 9 iterations for the flexible nozzles, the rate of improvement decelerates and the HV curves begin to plateau. This saturation behavior suggests that further iterations provide only marginal gains, indicating that the optimization process has effectively converged within the constrained design space. 

The slower convergence observed for the flexible nozzle can likely be attributed to the increased complexity of the underlying design space. Unlike the rigid nozzle, the performance of the flexible configuration is governed by strongly coupled FSI dynamics, which introduce additional nonlinearities into the relationship between nozzle geometry and the optimization objectives. Consequently, the optimizer requires a larger number of evaluations to adequately characterize the design space and identify high-performing regions before convergence is achieved.

\begin{figure}
    \centering
    \includegraphics[width=\linewidth]{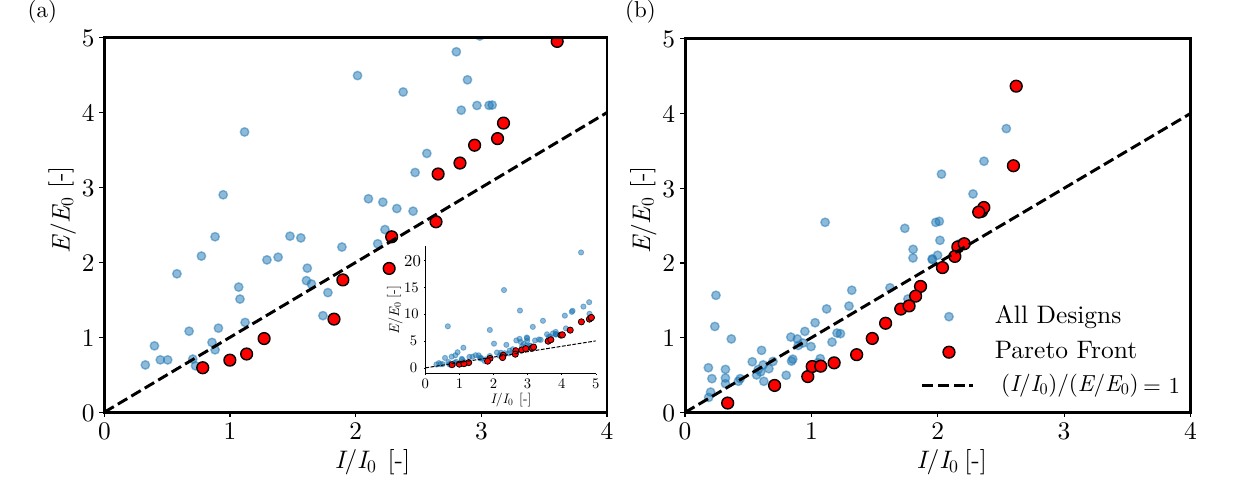}
    \caption{Pareto fronts obtained from MOBO of (a) rigid and (b) flexible nozzle configurations. Blue markers denote all evaluated designs, while red markers indicate Pareto-optimal solutions. Both nozzle types exhibit a trade-off between impulse enhancement and jet energy. However, the rigid nozzle achieves larger maximum impulse amplification, whereas the flexible nozzle attains comparable impulse levels at lower energy expenditure across the Pareto front.}
    \label{fig:paretoPlots}
\end{figure}

The resulting Pareto fronts obtained by MOBO are presented in Fig. \ref{fig:paretoPlots} (also refer to supplementary movies S1 and S2). Blue markers represent all evaluated designs, while Pareto-optimal solutions are highlighted in red for the rigid (Fig. \ref{fig:paretoPlots}(a)) and flexible (Fig. \ref{fig:paretoPlots}(b)) nozzle configurations. In total, 100 rigid nozzle designs and 80 flexible nozzle designs were evaluated during the optimization process. For both nozzle configurations, an increase in impulse is accompanied by a corresponding increase in jet energy, illustrating the inherent correlation between propulsive performance and energy cost. However, the shape and extent of the Pareto fronts differ substantially between the rigid and flexible cases.

The rigid nozzles exhibit a broader distribution of propulsive performance, extending to higher impulse values ($I/I_0 \approx 5$), but at the same time these are associated with substantially larger input energy requirements, as indicated by the steep rise in $E/E_0$. Here, $I$ and $E$ denote the impulse and energy associated with a given nozzle design, while $I_0$ and $E_0$ correspond to the baseline cylindrical nozzle. In contrast, the flexible nozzle exhibits a more compact Pareto front with a lower maximum impulse of approximately $I/I_0 \approx 2.5$, but achieves these performance levels at considerably lower energy costs.

Notably, for moderate impulse levels ($I/I_0 \lesssim 2$), the flexible Pareto front consistently lies below that of the rigid nozzle, indicating that comparable impulse can be generated with reduced energy input. This behavior suggests that structural flexibility improves propulsion efficiency by enabling more effective conversion of input energy into jet momentum, resulting in a more favorable impulse-energy trade-off. However, the maximum achievable 
impulse of the flexible nozzles remains lower than that of the rigid designs under the constraints of input energy. This is attributed to the deformation-induced modulation of the 
nozzle geometry, where pressure-driven radial expansion increases the effective outlet area during the pulse cycle in converging nozzle designs and reduces the ability to sustain highly 
accelerated jet exit conditions achievable with rigid geometries.

\begin{figure}[h!]
    \centering
    \includegraphics[width=\linewidth]{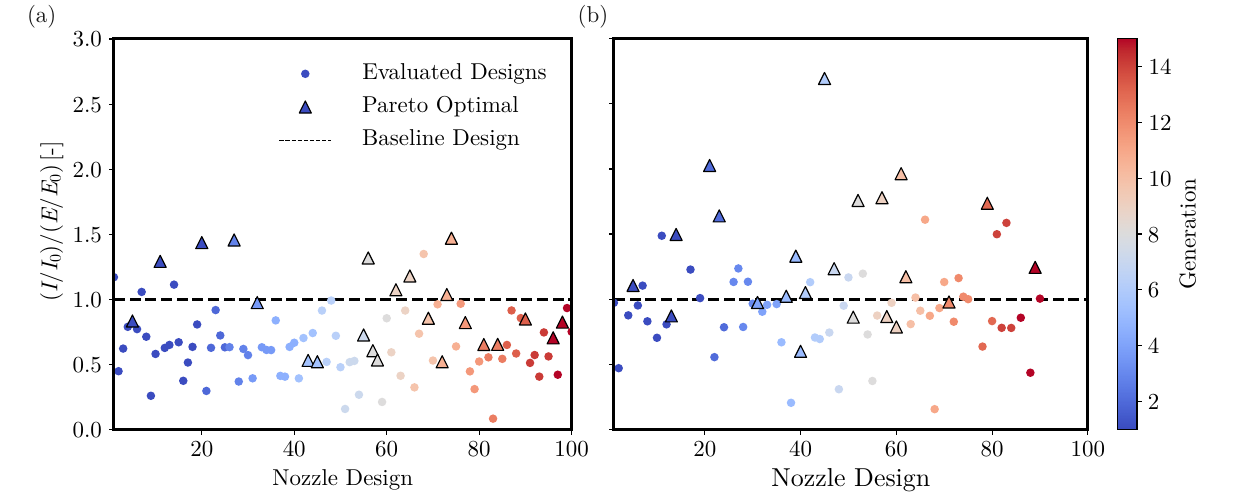}
    \caption{Normalized impulse-to-energy ratio, $(I/I_0)/(E/E_0)$, for all evaluated (a) rigid and (b) flexible nozzle designs. Color map denotes MOBO generation and triangles indicate Pareto-optimal designs. The dashed line corresponds to $(I/I_0)/(E/E_0)=1$, above which designs provide a net improvement in propulsive performance relative to the baseline cylindrical nozzle. Flexible nozzles achieve both a larger number of high-performing designs and a substantially higher maximum impulse-to-energy ratio than rigid nozzles.}
    \label{fig:paretoPlots_2}
\end{figure}

To further quantify this performance trade-off, Fig. \ref{fig:paretoPlots_2} evaluates the relative propulsive efficiency of the sampled designs, defined as the ratio of normalized impulse to normalized energy, $(I/I_0)/(E/E_0)$. Values greater than 1.0 (indicated by the dashed horizontal baseline) correspond to designs that generate a proportionally larger increase in impulse than in energy relative to the baseline cylindrical nozzle. Consequently, these designs provide an overall improvement in propulsion effectiveness compared to the reference configuration.

The Pareto-optimal solutions, indicated by triangles, consistently define the upper envelope of the design space in terms of the normalized impulse-to-energy ratio. The color mapping denotes the MOBO generation and reveals a clear progression toward higher-performing designs as the optimization proceeds. Early generations are broadly distributed throughout the design space, whereas later generations become increasingly concentrated near the upper boundary, demonstrating the ability of the MOBO framework to efficiently identify regions of improved performance. The relatively large performance variation observed across generations reflects the exploration component of the MOBO algorithm, which periodically samples regions of the design space associated with high predictive uncertainty in order to improve the accuracy of the surrogate model and avoid premature convergence to local optima.

A comparison between the rigid and flexible nozzle configurations reveals a significant advantage associated with structural flexibility. Among the 100 rigid-nozzle designs evaluated, only 8 Pareto-optimal designs satisfy $(I/I_0)/(E/E_0) > 1$, whereas 15 of the 80 flexible-nozzle designs exceed this threshold. Moreover, the maximum normalized impulse-to-energy ratio achieved by the flexible nozzle is approximately 2.7, substantially higher than the peak value of approximately 1.5 obtained for the rigid nozzle. These results indicate that flexibility shifts a larger fraction of the design space into a regime where impulse gains outweigh the associated increase in energy, resulting in a more favorable propulsion trade-off.

\begin{figure}[h!]
    \centering
    \includegraphics[width=\linewidth]{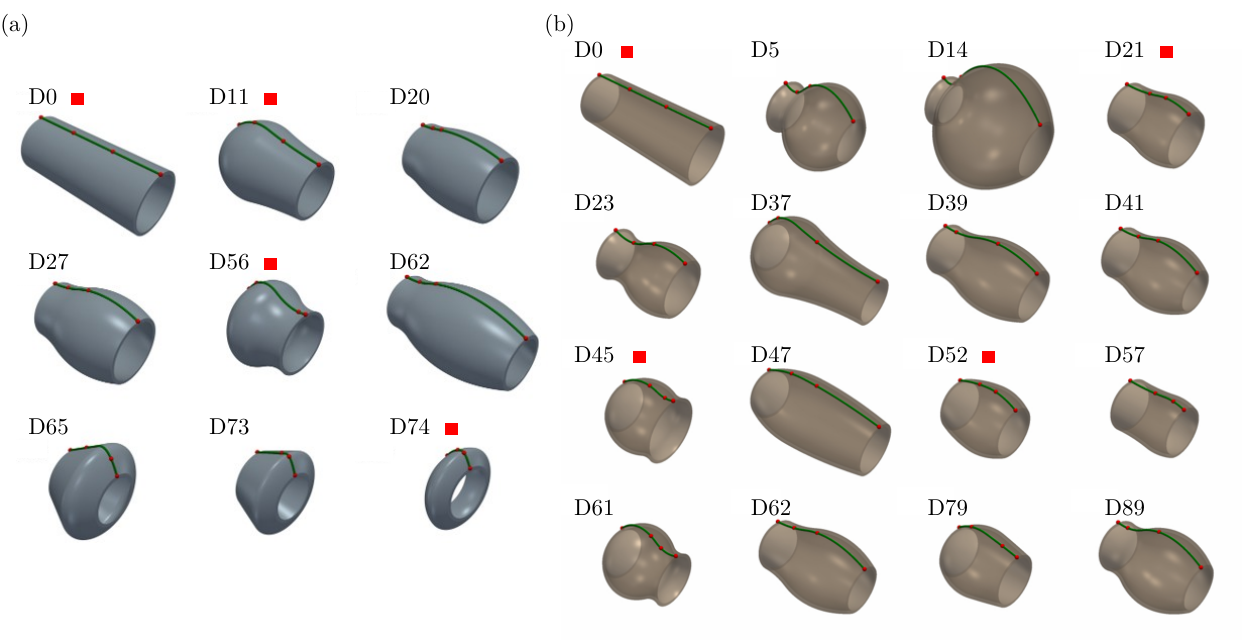}
    \caption{Pareto-optimal (a) rigid and (b) flexible nozzle geometries rendered in perspective view corresponding to designs satisfying $(I/I_0)/(E/E_0) > 1$, together with the baseline cylindrical nozzle (D0). The displayed configurations represent nozzle designs that achieve a proportionally greater increase in impulse than in energy relative to the baseline case. (Designs with red square marker are chosen for detailed physics analysis in subsequent sections.)}
    \label{fig:pareto_optimal_designs}
\end{figure}

To gain insight into the geometric characteristics associated with these high-performing designs, the nozzle profiles corresponding to all Pareto-optimal configurations satisfying $(I/I_0)/(E/E_0) > 1$ are presented in Fig. \ref{fig:pareto_optimal_designs}, together with the baseline cylindrical nozzle (Design 0 or D0). A wide variation in the optimized nozzle geometries can be observed, indicating that multiple shape configurations are capable of achieving improved propulsive performance. The influence of these geometric variations on the underlying flow and FSI mechanisms is examined in the following sections.

\CB{\subsection{Vortex Dynamics and Underlying Mechanisms of Pareto-Optimal Nozzles}}
\label{subsec:vortex_analysis}

To analyze the underlying flow mechanisms responsible for the improvements in propulsive performance, we select three of the best-performing rigid and flexible nozzle designs based on the relative efficiency metric $(I/I_0)/(E/E_0)$ and compare them with the baseline cylindrical nozzle (Design 0).\\

\subsubsection{Rigid Nozzles}

\begin{figure}[htbp]
    \centering
    \includegraphics[width=\linewidth]{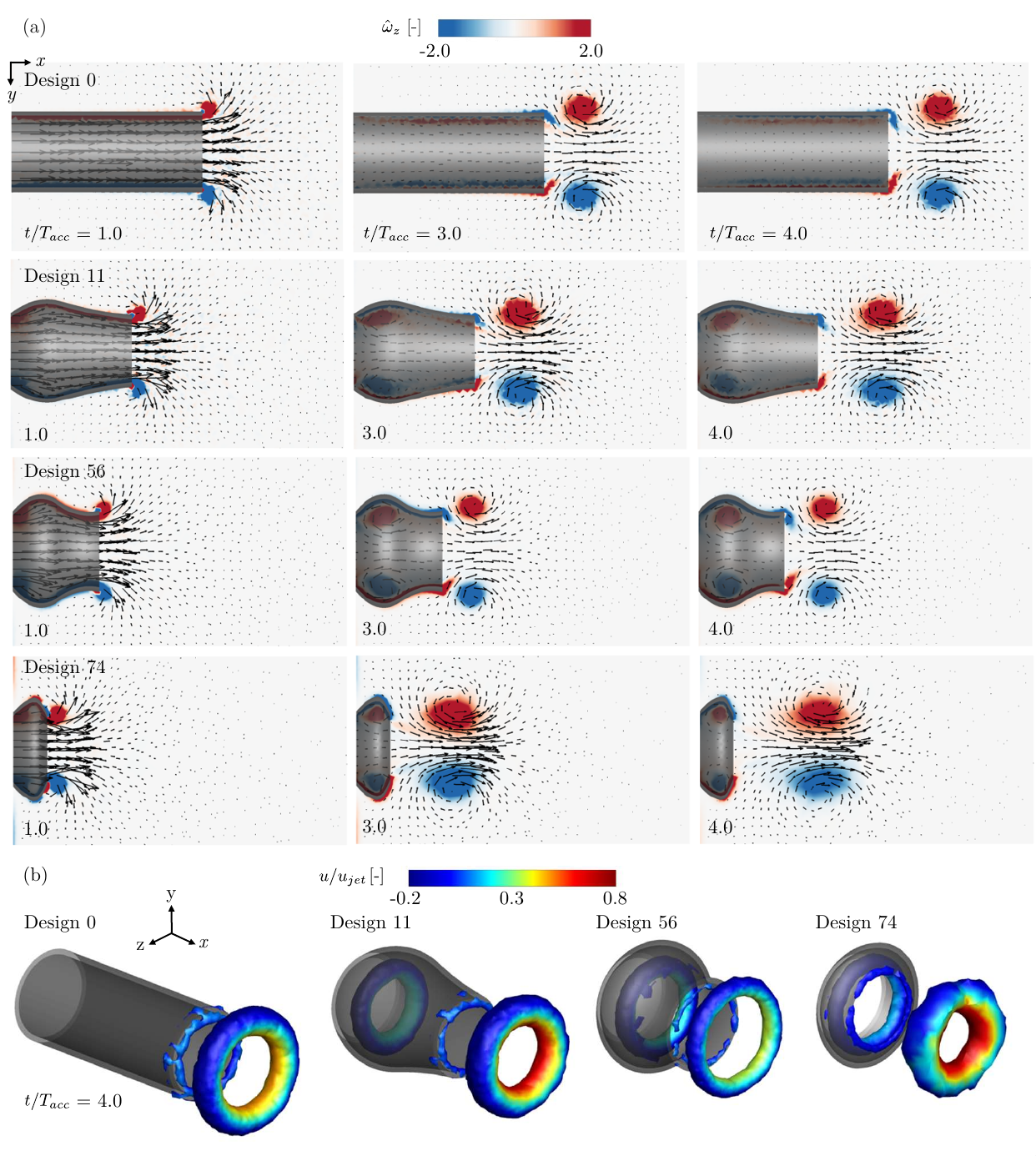}
    \caption{(a) Evolution of normalized out-of-plane vorticity contours $(\hat{\omega}_z)$ with velocity vectors on the central $x$-$y$ plane for the baseline and Pareto-optimal rigid nozzle designs at different phases of the pulsed jet cycle. (b) Iso-surfaces of $Q$-criterion ($Q = 200$) colored by the normalized axial velocity illustrating the three-dimensional vortex structures at $t/T_{\mathrm{acc}} = 4$.}
    \label{fig:rigid_vortex_contour}
\end{figure}

For the rigid nozzles, we select Designs 11, 56, and 74 with $(I/I_0)/(E/E_0)$ values of 1.29, 1.32, and 1.47, respectively. The influence of the optimized rigid geometries on the jet dynamics is investigated through the out-of-plane normalized vorticity contours, $\hat{\omega}_z = \omega_z D/u_{jet}$, with velocity vectors on the central $x$-$y$ cross-sectional plane shown in Fig.~\ref{fig:rigid_vortex_contour}(a), and the $Q$-criterion iso-surfaces ($Q = 200$) colored by the normalized axial velocity shown in Fig.~\ref{fig:rigid_vortex_contour}(b).

In the baseline cylindrical nozzle (Design 0), the pulsed jet undergoes nearly uniform axial acceleration, resulting in a conventional shear-layer roll-up followed by the formation of a symmetric counter-rotating vortex pair that pinches off and convects downstream during the jet deceleration phase. In contrast, the Pareto-optimal rigid nozzles exhibit substantially modified vortex dynamics and internal pressure distribution due to the axial shaping of the nozzle profile. A common feature observed across all optimized rigid geometries is the formation of a secondary vortex ring inside the nozzle prior to jet ejection, which is absent in the cylindrical baseline case (refer to supplementary movie S3). The diverging sections of the optimized geometries promote stronger vortex-induced internal entrainment within the nozzle, increasing the amount of fluid accelerated by the jet, while the subsequent converging sections further accelerate this entrained fluid prior to exit. The inlet diverging sections also reduce the pressure work required to drive the pulsed jet shown later in Section \ref{subsec:impulse_energy}. The combined effects of additional internal entrainment, downstream acceleration, and reduced pressure-energy expenditure increase axial momentum transfer and improve the conversion of input energy into useful propulsive impulse.

While this underlying mechanism is common across the selected Pareto-optimal designs, the extent of geometric curvature at the diverging-converging section influences the secondary vortex ring and the propulsive performance. Designs 11 and 74 amplify the primary vortex ring. Design 11 utilizes a gradual divergence and contraction section to smoothly accelerate the shear-layer roll-up, allowing the primary vortex pair to convect farther downstream than the baseline while maintaining a higher vorticity magnitude. Design 74 achieves the strongest vortex amplification through aggressive geometric shaping characterized by sharp diverging and converging sections. The sharp converging section resembles the orifice shape, in which the nozzle wall is orthogonal to the jet direction. This shape promotes the radial entrainment of ambient fluid, which increases the exit pressure (quantified later in Section \ref{subsec:impulse_energy}) and thereby increases overall thrust generation \cite{krieg2015pressure, limbourg2021formation}. Consequently, at $t/T_{\mathrm{acc}} = 4$, the vortex structures of Design 74 convect significantly farther downstream than all other configurations, supported by a stronger induced velocity field and amplified three-dimensional vortex structures with enhanced axial jet velocity presented in Fig. \ref{fig:rigid_vortex_contour}(b).

In contrast, Design 56 exhibits a more pronounced contraction toward the outlet originating from the highly curved diverging section near the nozzle mid-section. The strong geometric curvature near the mid-section generates a substantially stronger secondary vortex ring inside the nozzle compared to the other designs. While this enhances internal fluid entrainment, it also extracts a larger portion of the flow energy from the primary jet structure. As a result, the circulation growth of the primary vortex saturates earlier, and the downstream vortex convection remains lower than Designs 11 and 74.

\begin{figure}[htbp]
    \centering
    \includegraphics[width=\linewidth]{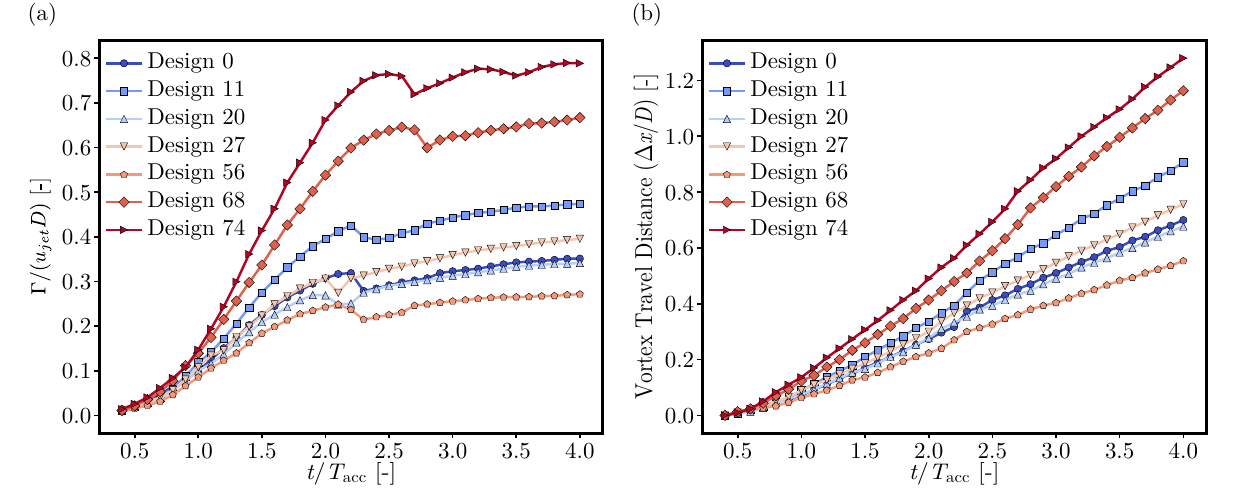}
    \caption{(a) Temporal evolution of the normalized primary vortex circulation $\Gamma$, for the baseline and Pareto-optimal rigid nozzle designs. (b) Temporal evolution of the axial displacement of the primary vortex core, $\Delta x$, from the nozzle outlet during the pulsed jet cycle, illustrating the enhanced vortex strength and convection in the optimized nozzle geometries.}
    \label{fig:rigid_vortex_quantification}
\end{figure}

To support the qualitative data in Fig. \ref{fig:rigid_vortex_contour}, the temporal evolution of the vortex circulation $\Gamma$ and the axial displacement of the primary vortex core from the nozzle outlet, $\Delta x$, are quantified in Fig. \ref{fig:rigid_vortex_quantification}. Here, the circulation is defined as $\Gamma = \iint_S \omega_z \, dA$ and is normalized using the peak jet velocity and nozzle diameter. As shown in Fig. \ref{fig:rigid_vortex_quantification}(a) and (b), most of the Pareto-optimal rigid nozzles generate larger circulation, faster vortex convection, and greater downstream displacement of the primary vortex than the baseline design throughout the pulsed jet cycle. This is consistent with the amplified vortex structures observed in the vorticity contours presented in Fig. \ref{fig:rigid_vortex_contour}. Design 74 exhibits the largest circulation growth and vortex-core displacement, reaching approximately 2.24 times and 1.83 times the baseline values, respectively, by the end of the pulsed jet cycle. By contrast, the energy extraction trade-off caused by the secondary vortex ring is clearly evident in a few cases, such as Design 56. Its primary vortex circulation and travel distance saturate early and remain lower than both the baseline and the other optimized designs. Overall, the enhanced vortex circulation and downstream convection observed in majority of the Pareto-optimal rigid nozzle designs demonstrate the direct relationship between optimized nozzle geometry, amplified vortex dynamics, and improved impulse-generation performance.

\subsubsection{Flexible Nozzles}

\begin{figure}[htbp]
    \centering
    \includegraphics[width=\linewidth]{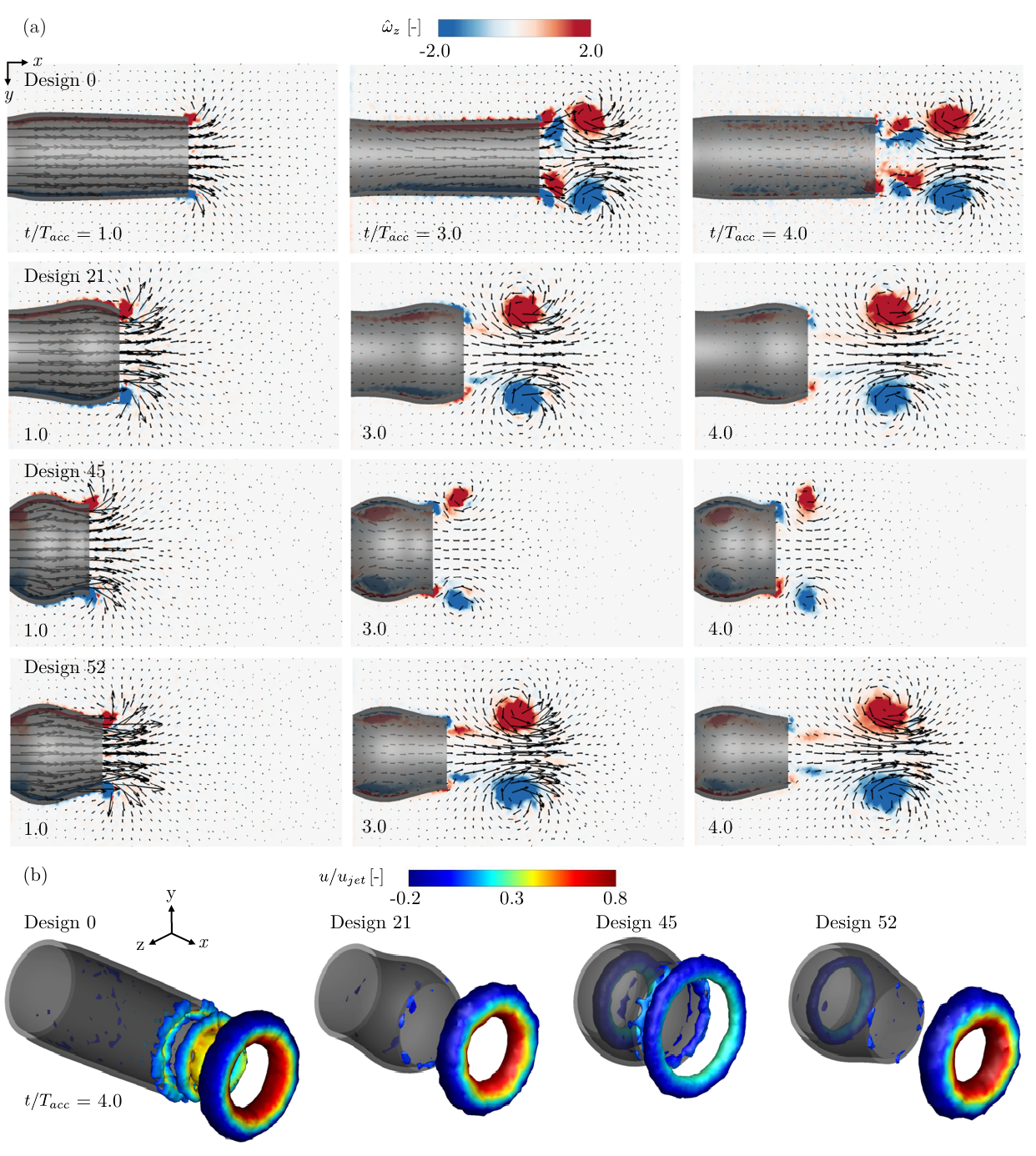}
    \caption{(a) Evolution of normalized out-of-plane vorticity contours $(\hat{\omega}_z)$ with velocity vectors on the central $x$-$y$ plane for the baseline and Pareto-optimal flexible nozzle designs at different phases of the pulsed jet cycle, illustrating the effect of FSI on vortex formation, jet entrainment, and downstream vortex convection. (b) Iso-surfaces of $Q$-criterion ($Q = 200$) colored by the normalized axial velocity illustrating the three-dimensional vortex structures at $t/T_{\mathrm{acc}} = 4$.}
    \label{fig:flexible_vortex_contour}
\end{figure}

Similar to the rigid nozzles, the vortex dynamics of the flexible nozzles are analyzed using the normalized vorticity contours and velocity-vector fields shown in Fig. \ref{fig:flexible_vortex_contour}(a), along with the corresponding three-dimensional vortex structures visualized using the $Q$-criterion iso-surfaces in Fig. \ref{fig:flexible_vortex_contour}(b). For the Pareto-optimal flexible nozzle configurations, we focus primarily on Designs 21, 45, and 52 with $(I/I_0)/(E/E_0)$ values of 2.03, 2.69, and 1.76, respectively.

\begin{figure}[htbp]
    \centering
    \includegraphics[width=\linewidth]{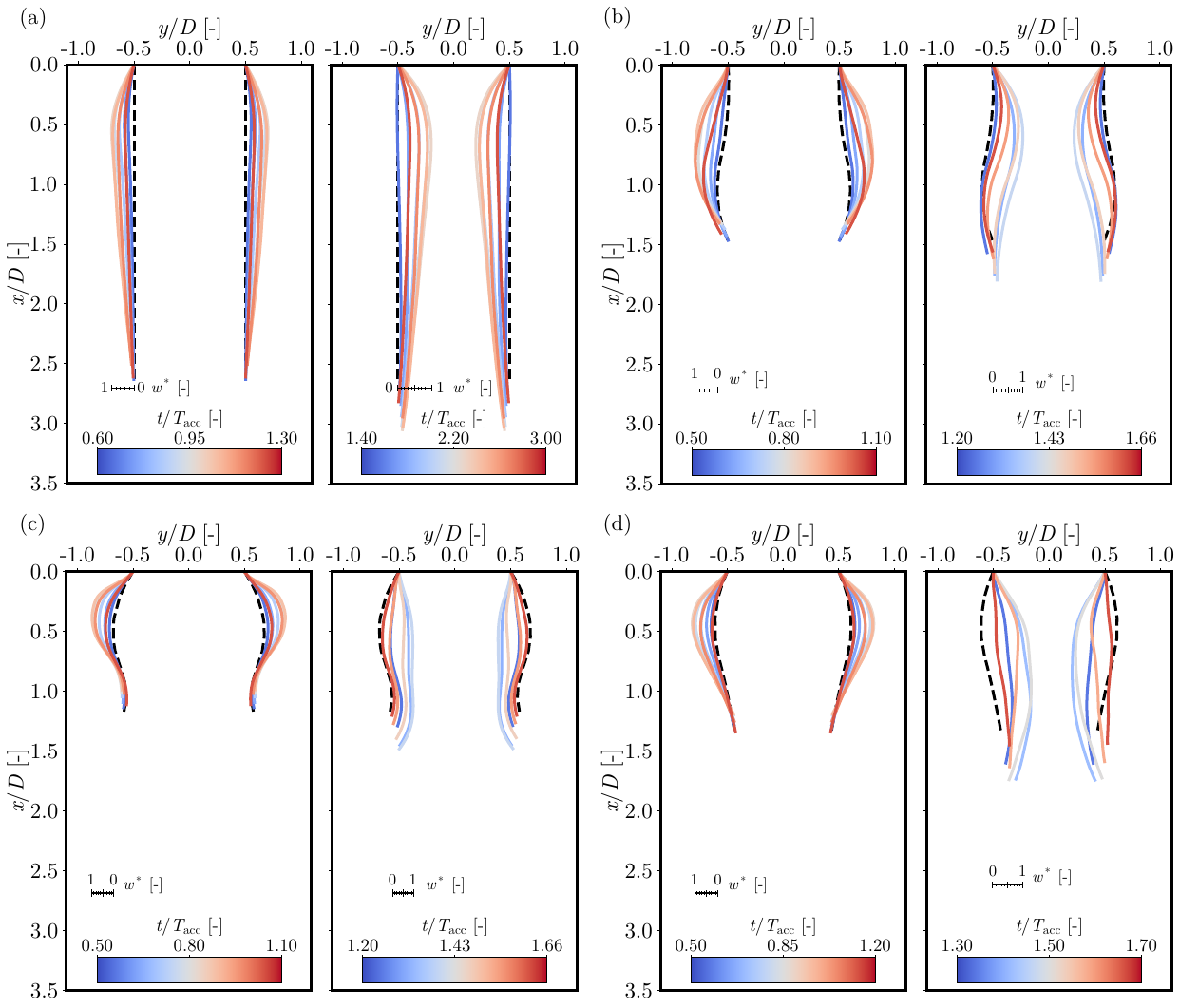}
    \caption{Deflection envelopes of the baseline (a) Design 0 and Pareto-optimal flexible nozzle designs: (b) Design 21, (c) Design 45, and (d) Design 52 during the expansion and contraction phases of the pulsed jet cycle. The dashed black lines represent the undeformed nozzle geometries, while the colored profiles denote the instantaneous deformed configurations. For visualization, the radial deformations are scaled such that the peak deformation amplitude extends by $0.2D$ from the undeformed configuration.}
    \label{fig:wave_deflection}
\end{figure}

Compared to the rigid cylindrical nozzle, the flexible nozzle geometries exhibit variation in vortex dynamics due to the FSI effects between the pulsatile internal flow and the deforming nozzle wall. To understand this underlying mechanism, Fig. \ref{fig:wave_deflection} presents the deformation envelopes of the nozzle walls during the pulsed jet cycle, where radial deformations are scaled and normalized such that the maximum amplitude extends by $0.2D$ from the undeformed configuration ($w^* = w/w_{\mathrm{max}}$). During the initial jet acceleration phase, the internal pressure rise drives a radial expansion wave that initiates near the nozzle inlet and propagates downstream towards the outlet. This expansion actively enhances internal fluid entrainment (often forming a secondary internal vortex) while temporarily delaying the roll-up of the primary shear layer at the exit and temporarily delaying the primary vortex formation process. Subsequently, the deformation wave reflects from the free end, initiating a rapid contraction phase. This traveling wave behavior results in multiple expansion-contraction cycles with wave speeds that depend on the nozzle geometry and deformation characteristics. As the nozzle contracts, the internally entrained fluid is expelled and accelerated downstream. This synchronized expansion-contraction dynamics acts as an active pumping mechanism (refer to supplementary movies S4 and S5), directly driving the enhanced impulse-generation efficiency observed in the flexible geometries.

Similar to rigid nozzles, the Pareto-optimal flexible nozzle designs containing an inlet diverging section form a secondary vortex structure within the nozzle that promotes vortex-induced entrainment in the nozzle interior. However, unlike the rigid nozzles where this entrainment mechanism is governed primarily by the static geometric shaping, the flexible nozzles additionally generate entrainment through time-dependent wall deformation as discussed earlier. As the nozzle contracts, the entrained fluid is expelled and accelerated downstream through the converging section, resulting in enhanced axial momentum transfer into the ambient fluid and consequently improved impulse-generation efficiency.

Among the selected configurations, Design 45 exhibits comparatively weaker primary vortex structures and reduced downstream convection despite achieving the highest efficiency metric among the selected flexible designs. The strong localized deformation near the nozzle mid-section generates a pronounced secondary vortex ring inside the nozzle, which redistributes a significant portion of the flow energy away from the primary vortex structure. Consequently, the growth of the primary vortex circulation saturates earlier, and the downstream convection remains substantially lower than the other designs. Nevertheless, the enhanced internal entrainment and contraction-driven acceleration of the flow still contribute to a significant improvement in the overall propulsive efficiency.

In contrast, Designs 21 and 52 exhibit substantially stronger vortex amplification and downstream convection. The larger deformation amplitudes and the resulting nozzle contraction generate stronger entrainment during the expansion phase while efficiently redirecting the entrained fluid downstream during contraction. At $t/T_{\mathrm{acc}} = 4$, these designs show stronger vortex pairs that convect significantly farther downstream than both the cylindrical baseline and Design 45. The corresponding velocity-vector fields further reveal stronger induced flow surrounding the vortex cores, indicating enhanced entrainment and momentum transport into the ambient fluid. The three-dimensional $Q$-criterion iso-surfaces shown in Fig. \ref{fig:flexible_vortex_contour}(b) additionally confirm the formation of amplified vortex-ring structures with elevated axial jet velocity, consistent with the enhanced impulse-generation performance of these flexible nozzle designs.

\begin{figure}[htbp]
    \centering
    \includegraphics[width=\linewidth]{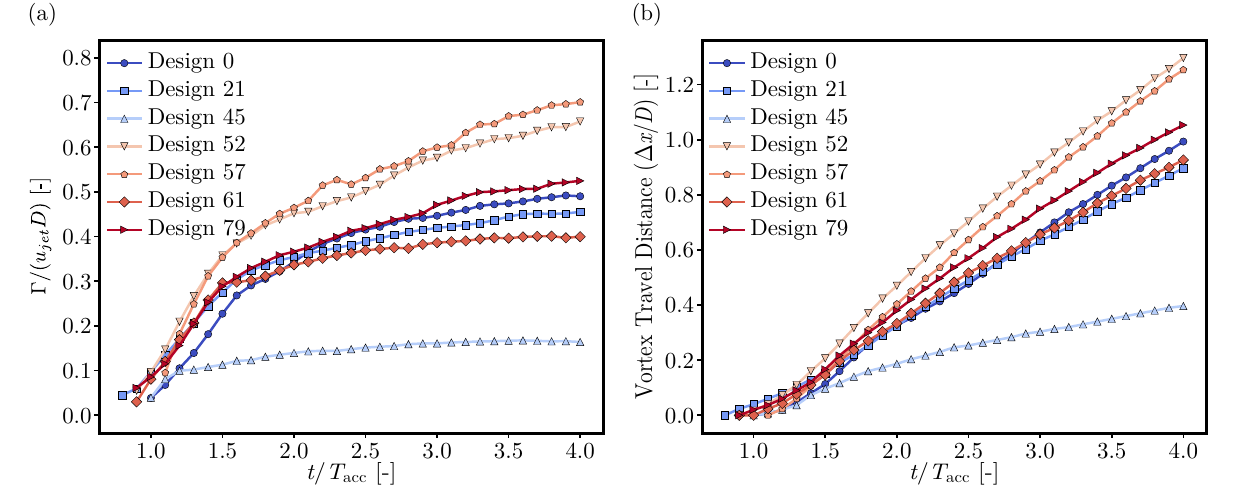}
    \caption{(a) Temporal evolution of the normalized primary vortex circulation, $\Gamma$, for the baseline and Pareto-optimal flexible nozzle designs. (b) Temporal evolution of the axial displacement of the primary vortex core, $\Delta x$, from the nozzle outlet during the pulsed jet cycle, illustrating the enhanced vortex amplification and downstream convection induced by FSI dynamics.}
    \label{fig:flexible_vortex_quantification}
\end{figure}

The temporal evolution of the primary vortex circulation and axial displacement of the vortex core from the nozzle outlet are quantified in Fig. \ref{fig:flexible_vortex_quantification}. Fig. \ref{fig:flexible_vortex_quantification}(a) presents the normalized circulation evolution for the selected flexible nozzle designs. Compared to the baseline cylindrical nozzle, most flexible configurations exhibit enhanced circulation growth throughout the pulsed jet cycle. In particular, Designs 52 and 57 demonstrate a rapid increase in circulation after $t/T_{\mathrm{acc}} \approx 1.0$, corresponding to the onset of the nozzle contraction phase following peak radial expansion. Among the presented cases, Design 57 exhibits the highest circulation growth, reaching approximately 1.43 times the circulation of the baseline cylindrical nozzle by the end of the pulsed jet cycle, followed closely by Design 52, which reaches approximately 1.34 times the baseline value.

A similar trend is observed in the axial displacement of the primary vortex cores shown in Fig. \ref{fig:flexible_vortex_quantification}(b). The high-performing flexible nozzle designs exhibit substantially faster downstream convection than the cylindrical baseline, indicating stronger axial momentum transfer into the ambient fluid. Design 52 produces the largest vortex-core displacement, with a convection distance approximately 1.3 times larger than the baseline nozzle by the end of the pulsed jet cycle. In contrast, as discussed earlier, Design 45 exhibits significantly weaker downstream convection due to the formation of a strong secondary vortex structure inside the nozzle, which redistributes a larger portion of the flow energy away from the primary vortex ring and limits its downstream propagation. Overall, the enhanced vortex circulation and downstream convection observed in the Pareto-optimal flexible nozzle designs demonstrate the important role of FSI, jet entrainment, and nozzle contraction dynamics in improving the propulsive performance of pulsed jet-flexible nozzle systems.

\subsection{Impulse and Energy Analysis of Pareto-optimal nozzles}
\label{subsec:impulse_energy}

\begin{figure}[htbp]
    \centering
    \includegraphics[width=\linewidth]{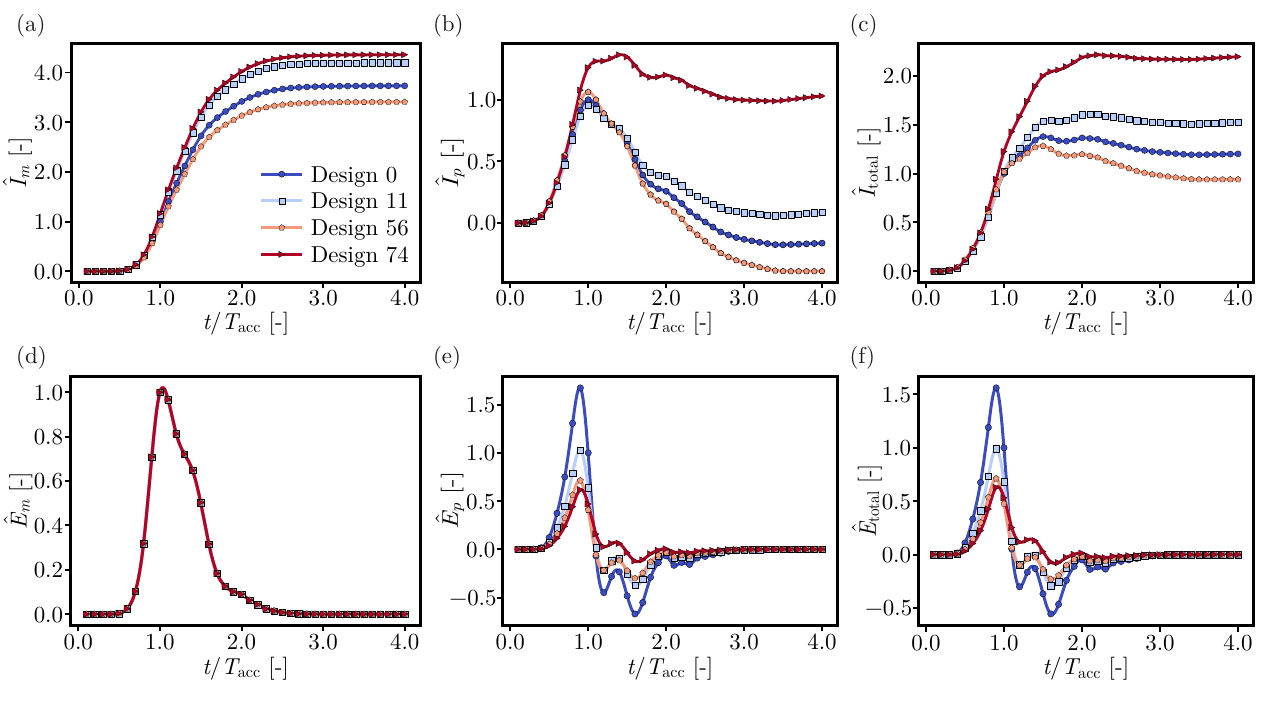}
    \caption{Temporal history of normalized impulse and jet energy input for the baseline and Pareto-optimal rigid nozzle designs during the pulsed jet cycle. Panels (a-c) show the normalized momentum impulse contribution $\hat{I}_m$, pressure impulse contribution $\hat{I}_p$, and total impulse $\hat{I}_{total}$, respectively. Panels (d-f) show the normalized kinetic energy contribution $\hat{E}_m$, pressure energy contribution $\hat{E}_p$, and total jet energy input $\hat{E}_{total}$, respectively. All quantities are normalized using the corresponding baseline cylindrical nozzle (Design 0) values at $t/T_{\mathrm{acc}} = 1$.}
    \label{fig:rigid_impulse_energy_history}
\end{figure}

\begin{figure}[htbp]
    \centering
    \includegraphics[width=\linewidth]{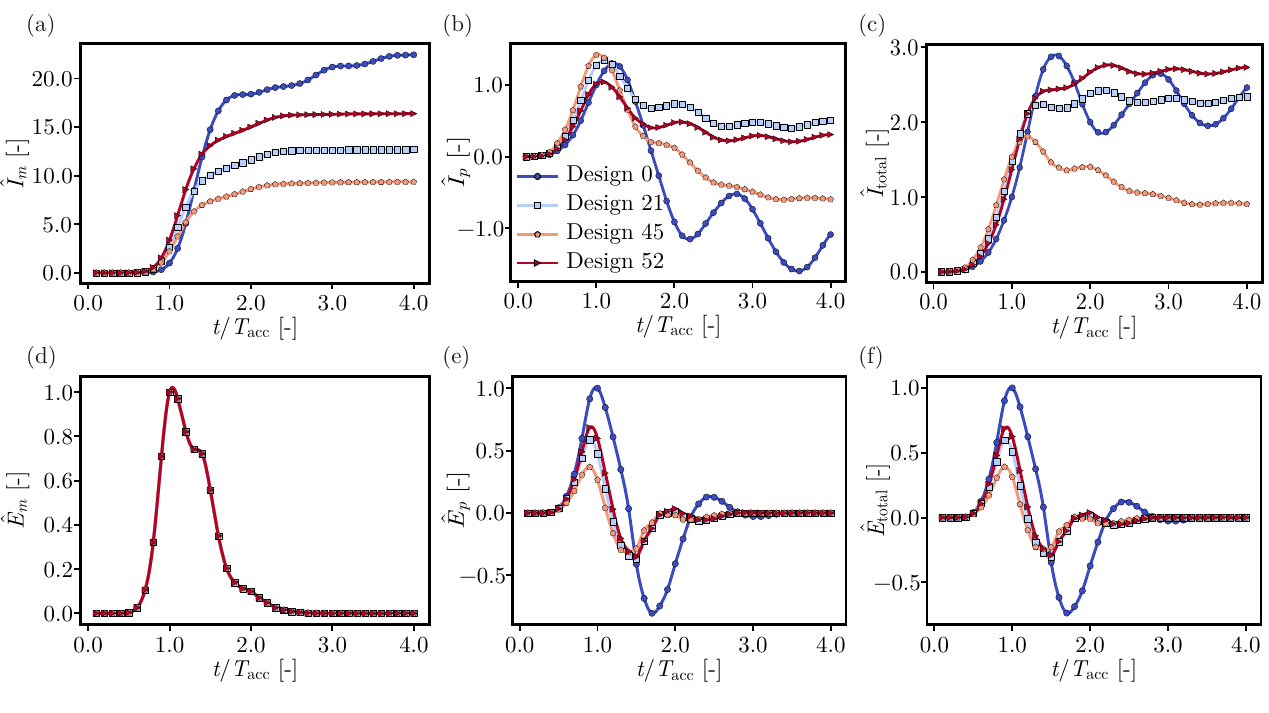}
    \caption{Temporal history of normalized impulse and jet energy input for the baseline and Pareto-optimal flexible nozzle designs during the pulsed jet cycle. Panels (a-c) show the normalized momentum impulse contribution $\hat{I}_m$, pressure impulse contribution $\hat{I}_p$, and total impulse $\hat{I}_{total}$, respectively. Panels (d-f) show the normalized kinetic energy contribution $\hat{E}_m$, pressure energy contribution $\hat{E}_p$, and total jet energy input $\hat{E}_{total}$, respectively. All quantities are normalized using the corresponding baseline cylindrical nozzle (Design 0) values at $t/T_{\mathrm{acc}} = 1$.}
    \label{fig:flexible_impulse_energy_history}
\end{figure}

In Section \ref{subsec:convergence}, we presented the time-integrated impulse and energy values for the nozzle designs throughout the optimization process. To further illustrate the propulsive performance of the Pareto-optimal nozzle designs, the temporal evolution of impulse and jet energy input during the pulsed jet cycle is analyzed for both the rigid and flexible nozzle configurations. Fig. \ref{fig:rigid_impulse_energy_history} and \ref{fig:flexible_impulse_energy_history} present the corresponding impulse and energy histories for the Pareto-optimal rigid and flexible nozzle designs, respectively. For both cases, panels (a-c) show the normalized momentum impulse contribution $\hat{I}_m$, pressure impulse contribution $\hat{I}_p$, and total impulse $\hat{I}_{total}$, while panels (d-f) show the normalized kinetic energy contribution $\hat{E}_m$, pressure energy contribution $\hat{E}_p$, and total jet energy input $\hat{E}_{total}$. All quantities are normalized using the corresponding baseline cylindrical nozzle (Design 0) values at $t/T_{\mathrm{acc}} = 1$.

Fig. \ref{fig:rigid_impulse_energy_history}(a) shows that the optimized rigid nozzle geometries generally produce larger momentum impulse than the baseline cylindrical nozzle during the pulsed jet cycle. The increase in $\hat{I}_m$ is consistent with the enhanced vortex circulation and downstream convection discussed earlier in Section \ref{subsec:vortex_analysis}. In particular, Designs 11 and 74 exhibit significantly larger momentum impulse growth due to the increased internal entrainment and geometric contraction-induced acceleration generated by the optimized profiles. Among all the rigid configurations, Design 74 achieves the highest momentum impulse, reaching approximately 1.17 times larger values than the baseline nozzle by the end of the pulsed jet cycle. In contrast, Design 56 produces comparatively lower momentum impulse growth despite its improved propulsive efficiency. As discussed earlier, the stronger secondary vortex structure generated inside the nozzle redistributes a larger fraction of the flow energy away from the primary jet momentum, thereby limiting the overall increase in outlet momentum.

It can be observed from the pressure impulse contribution in Fig. \ref{fig:rigid_impulse_energy_history}(b) that during the early stages of the pulsed jet cycle, all nozzle designs experience a rapid increase in $\hat{I}_p$ associated with the pressure rise required to accelerate the fluid through the nozzle. However, as vortex formation and downstream convection intensify, the pressure contribution decreases and eventually becomes negative for several configurations due to the acceleration of the jet and the associated reduction in outlet pressure. Design 74 maintains substantially higher pressure impulse throughout the pulse cycle, indicating stronger pressure-driven acceleration and enhanced momentum transfer at the nozzle exit.

The combined effect of the momentum and pressure contributions is reflected in the total impulse evolution shown in Fig. \ref{fig:rigid_impulse_energy_history}(c). Since the increase in momentum impulse dominates over the reduction in pressure contribution during the later stages of the pulse cycle, the Pareto-optimal rigid nozzles (D11 and D74) achieve higher total impulse than the baseline cylindrical configuration. Among the selected cases, Design 74 produces the largest total impulse, reaching approximately 1.83 times the baseline value by the end of the pulsed jet cycle.

The corresponding jet input energy characteristics are shown in Fig. \ref{fig:rigid_impulse_energy_history}(d-f). Fig. \ref{fig:rigid_impulse_energy_history}(d) shows that the kinetic energy contribution $\hat{E}_m$ remains identical for all rigid nozzle configurations because the inlet momentum forcing is prescribed identically for each case. In contrast, the pressure energy contribution $\hat{E}_p$ shown in Fig. \ref{fig:rigid_impulse_energy_history}(e) varies substantially among the nozzle geometries. The optimized nozzles generally exhibit lower pressure energy expenditure than the baseline cylindrical nozzle due to the diverging section close to the inlet, indicating more efficient conversion of the supplied energy into useful jet momentum. In particular, Design 74 achieves the lowest pressure energy contribution during the later stages of the pulse cycle while simultaneously generating the highest total impulse.

Consequently, the total normalized energy transfer $\hat{E}_{total}$ shown in Fig. \ref{fig:rigid_impulse_energy_history}(f) decreases for the optimized rigid nozzles relative to the baseline configuration. The reduction in total energy expenditure combined with the enhanced impulse generation explains the improved propulsive efficiency observed in the Pareto-optimal rigid nozzle designs. 

For the flexible nozzle configurations, the impulse histories shown in Fig. \ref{fig:flexible_impulse_energy_history}(a) exhibit fundamentally different characteristics compared to the rigid nozzle cases. Unlike the Pareto-optimal rigid nozzle designs, the selected flexible nozzles designs exhibit lower peak momentum impulse contributions, $\hat{I}_m$, than the baseline cylindrical nozzle. During the initial jet acceleration phase, the optimized flexible designs (D21, D45, and D52) exhibit higher momentum impulse, likely due to the faster wave propagation speeds leading to a more rapid expansion-contraction response in these comparatively shorter nozzles. However, during the later stages of the pulsed jet cycle corresponding to the nozzle contraction phase, the baseline cylindrical nozzle begins to show larger momentum impulse growth. This behavior indicates that, unlike the rigid nozzles where geometric shaping directly amplifies outlet jet momentum, the flexible nozzle dynamics redistribute a portion of the supplied energy into the wall deformation and secondary vortex formation.

The pressure impulse contribution shown in Fig. \ref{fig:flexible_impulse_energy_history}(b), however, show trends consistent with the optimized rigid nozzle designs. During the early stages of the pulse cycle, all flexible nozzle designs experience a rapid increase in $\hat{I}_p$ associated with the pressure rise required to drive the expanding flow through the deforming nozzle. However, as the contraction phase initiates and the jet accelerates downstream, the outlet pressure decreases, causing the pressure impulse contribution to reduce and become negative for designs 0 and 45. Compared to the baseline cylindrical nozzle, the Pareto-optimal flexible nozzles maintain substantially higher pressure impulse during the later stages of the pulse cycle, indicating more effective pressure recovery resulting from the FSI dynamics during the jet deceleration phase. 

The combined effect of the momentum and pressure contributions is reflected in the temporal evolution of the total impulse shown in Fig. \ref{fig:flexible_impulse_energy_history}(c). Unlike the rigid nozzles, where the momentum contribution dominates the impulse enhancement, the flexible nozzle performance is governed primarily by the mitigation of the negative pressure impulse during the later stages of the pulse cycle. Consequently, Designs 21 and 52 maintain relatively stable total impulse values after the initial growth phase, whereas the baseline cylindrical nozzle exhibits noticeable fluctuations in total impulse caused by the strong negative pressure impulse associated with its repeated expansion-contraction cycles.

By the end of the pulsed jet cycle, Design 52 achieves the highest total impulse among the selected configurations, reaching approximately 1.11 times the baseline value. Design 21 also maintains consistently higher total impulse throughout the pulse cycle, while Design 45 exhibits comparatively lower total impulse growth because of the earlier saturation of the primary vortex circulation discussed previously.

The corresponding jet energy input time histories are shown in Fig. \ref{fig:flexible_impulse_energy_history}(d-f). Similar to the rigid nozzle cases, the kinetic energy contribution $\hat{E}_m$ shown in Fig. \ref{fig:flexible_impulse_energy_history}(d) remains identical for all flexible nozzle configurations because the inlet momentum forcing is prescribed identically. In contrast, the pressure energy contribution $\hat{E}_p$ shown in Fig. \ref{fig:flexible_impulse_energy_history}(e) differs substantially among the nozzle designs due to the FSI effects and deformation-induced pressure variations. Compared to the cylindrical baseline nozzle, the Pareto-optimal flexible configurations exhibit significantly lower pressure energy expenditure during the acceleration and deceleration stages of the pulse cycle. 

Consequently, the total jet energy input shown in Fig. \ref{fig:flexible_impulse_energy_history}(f) is substantially reduced for the Pareto-optimal flexible nozzles relative to the baseline cylindrical configuration. This significant reduction in jet energy input, combined with the improved impulse characteristics, leads to the overall enhancement in propulsive efficiency observed in the flexible nozzle designs.

\section{Conclusions}
\label{sec:conclusions}

\CB{The present study proposes a multi-objective Bayesian optimization framework for designing energy-efficient rigid and flexible nozzles in pulsed jet-based propulsion systems. This study primarily focuses on maximizing impulse generation while minimizing jet energy input. Nozzle geometries are parameterized using a two-dimensional B-spline representation and modeled as axisymmetric to reduce the dimensionality of the optimization problem. By integrating three-dimensional FSI simulations with a Gaussian process surrogate model, the framework efficiently explore the nonlinear design space and identify Pareto-optimal configurations that balance the competing objectives. According to the hypervolume convergence metric, the optimizer identifies eight rigid and fifteen flexible Pareto-optimal nozzle designs that achieve relative propulsive efficiencies greater than unity, $(I/I_0)/(E/E_0) > 1$.}

\CB{Pareto-front analysis reveals distinct performance characteristics for rigid and flexible nozzle systems. Rigid nozzles achieve significantly greater impulse amplification, reaching up to five times the baseline cylindrical nozzle impulse, but these improvements require substantially higher jet energy input. In contrast, flexible nozzles produce lower peak impulse enhancements but demonstrate more favorable impulse-to-energy trade-offs. A larger proportion of the flexible-nozzle design space achieves $(I/I_0)/(E/E_0) > 1$, indicating that structural flexibility enables more efficient conversion of input energy into propulsive output. As a result, flexible nozzles attain a maximum normalized impulse-to-energy ratio of approximately 2.7, surpassing the peak value of approximately 1.5 observed for rigid nozzle configurations.}

\CB{Flow physics analysis indicates that performance improvements in rigid nozzles are primarily driven by geometry-induced modifications to vortex dynamics. Optimized diverging-converging nozzle profiles generate secondary vortex structures within the nozzle interior, enhancing internal entrainment and accelerating the entrained fluid prior to ejection. These mechanisms result in stronger vortex circulation, faster downstream vortex convection, and increased momentum transfer to the surrounding fluid, thereby augmenting impulse generation.}

\CB{In flexible nozzle configurations, propulsive mechanisms are governed by strongly coupled FSI dynamics. Deforming nozzle walls generate traveling expansion-contraction waves, which promote additional entrainment within the nozzle during expansion and accelerate the entrained fluid during contraction. Unlike rigid nozzles, where impulse enhancement is primarily due to increased outlet momentum, flexible nozzle designs improve propulsive efficiency mainly by reducing pressure-energy expenditure and mitigating unfavorable pressure impulse contributions during the later stages of the pulsed jet cycle. These FSI-driven mechanisms enable flexible nozzles to achieve substantial reductions in total jet energy input while maintaining enhanced impulse-generation characteristics.}

\CB{Impulse and energy analyses further highlight the distinct performance characteristics of the optimized nozzle systems. Rigid nozzles achieve increased total impulse through amplified momentum transport and stronger vortex convection, while flexible nozzles improve efficiency primarily via pressure-recovery mechanisms associated with nozzle deformation and wave propagation. In both cases, the optimized geometries reduce the total pressure energy required to drive the pulsed jet compared to the baseline cylindrical nozzle, demonstrating the optimization framework's effectiveness in identifying energy-efficient propulsive nozzle designs.}

\CB{In summary, this study demonstrates that nozzle geometry and structural flexibility are critical in governing vortex formation, entrainment, momentum transfer, pressure recovery, and energy conversion in pulsed jet propulsion systems. While rigid nozzle optimization maximizes absolute impulse generation, flexible nozzles offer a more energy-efficient propulsion strategy by leveraging FSI mechanisms to improve the impulse-to-energy trade-off. These findings offer new insights for the design of bio-inspired and energy-efficient pulsed-jet propulsion systems and establish a general computational framework for optimizing coupled fluid and structural dynamics in unsteady propulsion applications.}

\begin{acknowledgments}
This work was supported by the DARPA Young Faculty Award - DARPA-RA-24-01-18-YFA18-FP-004 (PI: Saad Bhamla) and the National Research Foundation of Korea - RS-2022-NR070924 (PI: Daehyun Choi). The present FSI simulations are carried out using the computational resources provided by the UK national supercomputing facility ARCHER2 through the EPSRC Access To HPC Pioneer Grant \CB{and UK Turbulence Consortium allocation} (PI: Chandan Bose). 
\end{acknowledgments}

\section*{Data Availability}
\CB{The data that support the findings of this study are available from the corresponding authors upon reasonable request.}

\appendix
\section{MOBO Computational Framework Validation}

To assess the accuracy and convergence characteristics of the proposed MOBO framework, validation studies were performed using four standard benchmark problems commonly employed in the multi-objective optimization literature: the Schaffer, Fonseca, Poloni, and Tanaka functions. These benchmark problems were selected because they exhibit a wide range of Pareto-front characteristics, including smooth convex fronts, highly nonlinear trade-offs, and disconnected Pareto-optimal regions. The mathematical definitions and design-variable bounds for each benchmark problem are summarized in Table \ref{tab:validation_benchmark}.

\begin{table}[htbp]
    \centering
    \caption{Benchmark functions used for MOBO validation, adapted from \cite{GALUZIO_MOBO}}
    \label{tab:validation_benchmark}
    \begin{tabular}{cccc}
    \hline
    Name & $n$ & Bounds & Functions \\
    \hline
    Schaffer & 1 & $[-10^3,\ 10^3]$ &
    $\begin{aligned}
        &f_1(x) = x^2 \\
        &f_2(x) = (x-2)^2
    \end{aligned}$ \\
    \hline
    Fonseca & 3 & $[-4,\ 4]$ &
    $\begin{aligned}
        &f_1(x) = 1 - \exp\!\left(-\sum_{i=1}^{3}\left(x_i - \dfrac{1}{\sqrt{3}}\right)^2\right) \\
        &f_2(x) = 1 - \exp\!\left(-\sum_{i=1}^{3}\left(x_i + \dfrac{1}{\sqrt{3}}\right)^2\right)
    \end{aligned}$ \\
    \hline
    Poloni & 2 & $[-\pi,\ \pi]$ &
    $\begin{aligned}
        &f_1(x) = \left[1 + (A_1-B_1)^2 + (A_2-B_2)^2\right] \\
        &f_2(x) = \left[(x_1+3)^2 + (x_2+1)^2\right] \\
        &A_1 = 0.5\sin 1 - 2\cos 2 + \sin 2 - 1.5\cos 2 \\
        &A_2 = 1.5\sin 1 - \cos 1 + 2\sin 2 - 0.5\cos 2 \\
        &B_1 = 0.5\sin x_1 - 2\cos x_1 + \sin x_2 - 1.5\cos x_2 \\
        &B_2 = 1.5\sin x_1 - \cos x_1 + 2\sin x_2 - 0.5\cos x_2
    \end{aligned}$ \\
    \hline
    Tanaka & 2 & $[0,\ \pi]$ &
    $\begin{aligned}
        &f_1(x) = x_1 \\
        &f_2(x) = x_2
    \end{aligned}$ \\
    \hline
    \end{tabular}
\end{table}

For each benchmark problem, the optimization was initialized using LHS, followed by sequential BO using GPR surrogate models and the q-batch Log Expected Hypervolume Improvement acquisition strategy employed throughout the present study. A total of $N=100$ objective-function evaluations were performed for each benchmark case. The Pareto-optimal solutions identified by the MOBO framework were then compared with the corresponding analytically known true Pareto fronts (TPFs).

\begin{figure}[htbp]
    \centering
    \includegraphics[width=\linewidth]{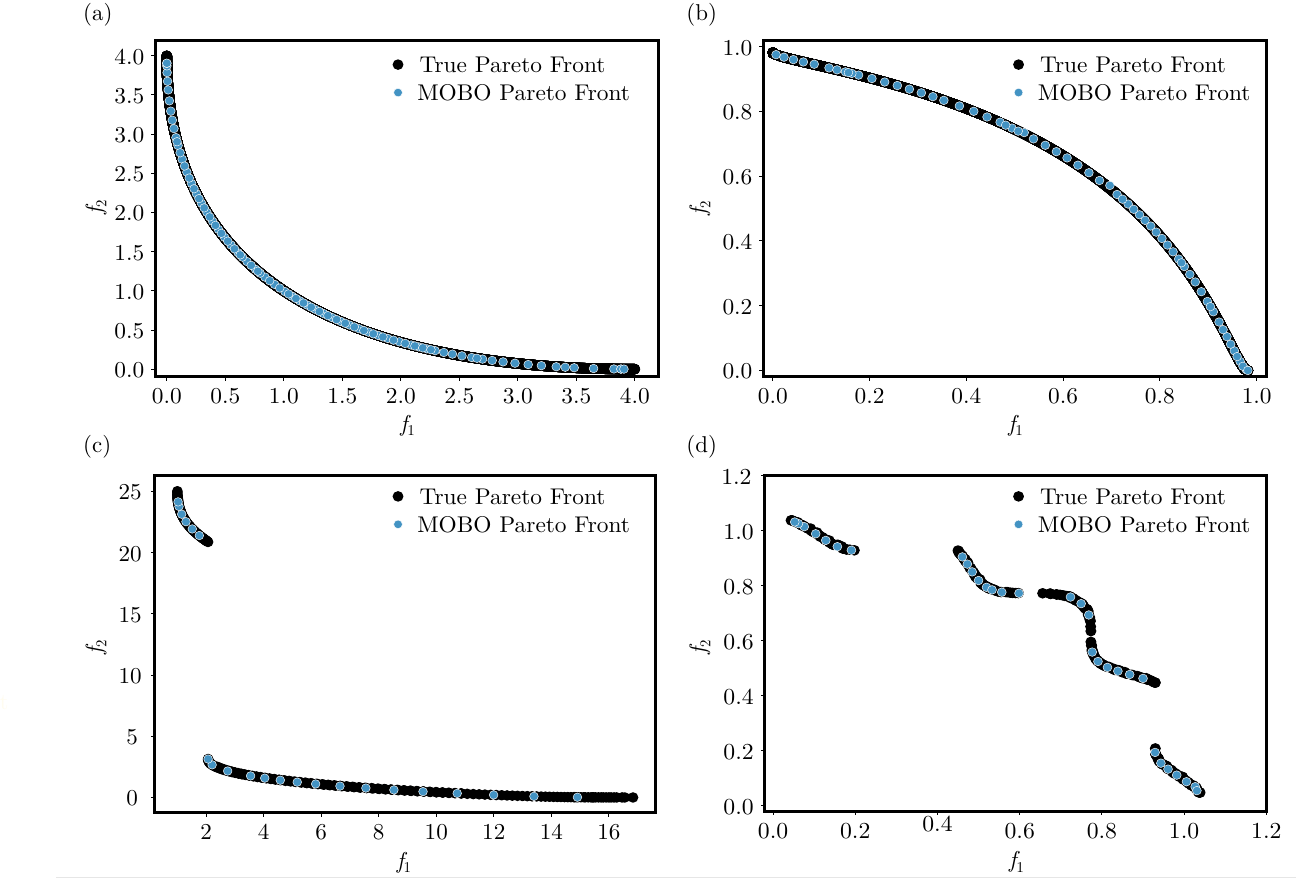}
    \caption{Validation of the MOBO framework against four standard multi-objective benchmark functions described in \ref{tab:validation_benchmark}: (a) Schaffer, (b) Fonseca, (c) Poloni, and (d) Tanaka. Blue markers show the Pareto front identified by MOBO after $N=100$ function evaluations; black markers show the analytically known TPFs. The close agreement across all four test cases, including the discontinuous fronts in the Poloni and Tanaka problems, demonstrates the ability of the GPR surrogate model-based MOBO framework to efficiently identify Pareto-optimal solutions with limited function evaluations.}
    \label{fig:MOBO_validation}
\end{figure}

The resulting Pareto fronts are shown in Fig. \ref{fig:MOBO_validation}. For all benchmark problems, the Pareto fronts predicted by the MOBO framework exhibit close agreement with the corresponding TPFs. In the Schaffer and Fonseca problems, the algorithm accurately reproduces the smooth continuous Pareto-front geometry over the entire objective space. Similarly, for the more challenging Poloni and Tanaka problems, which contain disconnected and highly nonlinear Pareto-optimal regions, the MOBO framework successfully identifies all major Pareto-front segments and accurately captures their shape and distribution.

Overall, the benchmark results demonstrate that the proposed GPR-based MOBO framework is capable of efficiently approximating complex Pareto fronts while requiring a relatively small number of objective-function evaluations. The close agreement between the predicted and analytical Pareto fronts provides confidence in the ability of the optimization framework to identify high-quality Pareto-optimal solutions for the computationally expensive CFD and FSI nozzle-optimization problems considered in the main body of this work.

\bibliography{apssamp}

\end{document}